# Sparse Sensor Placement for Reducing Forecast Errors in Ensemble Kalman Filtering

Takumi Saito[1, 2] and Shunji Kotsuki[3, 4]

[1] Graduate School of Science and Engineering, Chiba University, Chiba, Japan
[2] Japan Meteorological Agency, Tokyo, Japan
[3] Center for Environmental Remote Sensing, Chiba University, Chiba, Japan
[4] Institute for Advanced Academic Research, Chiba University, Chiba, Japan

**Correspondence:** Takumi Saito, Graduate School of Science and Engineering, Chiba University, Chiba, Japan.
**Email:** takumi_saito@chiba-u.jp

## Abstract

Designing efficient observation networks for reducing forecast errors is a fundamental challenge in numerical weather prediction. Data-driven sparse sensor placement (SSP) and ensemble-based data assimilation via the Ensemble Kalman Filter (EnKF) have each addressed this challenge independently, yet their mathematical connections have not been systematically formalized. This study presents a unified theoretical framework integrating SSP and EnKF through optimal experimental design, providing new theoretical and algorithmic results. While conventional SSP methods aim to reduce analysis errors, this study extends the SSP to target forecast error reduction by using a tangent linear model approximated by an ensemble forecast. In the context of data assimilation, the observation network design problem is optimization of observation locations (or observation operator) for reducing forecast errors. We derive the Fisher information matrices in the ensemble and model spaces for the EnKF, and clarify the mathematical interpretations of A-, D-, and E-optimality in terms of forecast error reduction. A-optimality in the model space minimizes the mean forecast error variance; D-optimality is ill-defined in the model space due to rank deficiency and is therefore formulated in the ensemble space, where it maximizes the Shannon information content of assimilated observations; and E-optimality in the model space minimizes the worst-case forecast error variance. We further propose a fast greedy algorithm for selecting observation locations under A-optimality in the model space, avoiding matrix inversion at each greedy step and substantially reducing computational cost. Numerical experiments using the Lorenz-96 model support these theoretical findings. Among the three optimality criteria, A-optimality in the model space most consistently reduces forecast spread and root-mean-square error, and yields stable incremental improvements consistent with post-assimilation observation impact diagnostics. These results indicate that A-optimality in the model space with the fast greedy algorithm is the most reliable and well-balanced approach for observation network design in ensemble-based data assimilation.

Keywords: sparse sensor placement, data assimilation, ensemble Kalman filter, adaptive observations, Fisher information matrix, optimal experimental design, greedy algorithm

## 1. Introduction

Large-scale dynamical systems in geosciences and engineering require observations at numerous locations to accurately estimate their high-dimensional states. In practice, however, observational resources are limited by physical, economic, and logistical constraints, making dense sensor deployment infeasible. This motivates the problem of observation network design: given a limited number of observations, how should they be positioned to maximize their impact on state estimation accuracy? This question arises across a broad range of applications, from structural health monitoring and flow diagnostics to atmospheric and oceanic state estimation.

Data-driven sparse sensor placement (SSP) has emerged as a principled approach to addressing this challenge by exploiting the low-dimensional structure inherent in high-dimensional datasets. In the SSP pioneered by Manohar et al. (2018), training data are subjected to proper orthogonal decomposition (POD) implemented via singular value decomposition (SVD) to extract a low-dimensional basis, and observation locations are selected to maximize the reconstruction accuracy of the low-dimensional state. This approach was subsequently interpreted as an optimal experimental design problem based on a Fisher information matrix (FIM) by Saito et al. (2021) and Nakai et al. (2021). The widely adopted optimality criteria — A-optimality, D-optimality, and E-optimality (Pukelsheim, 2006) — minimize the trace of the inverse of the FIM, maximize the determinant of the FIM, and maximize the minimum eigenvalue of the FIM, respectively. In the SSP context, A-optimality has been adopted to minimize mean squared error in state estimation (Nakai et al., 2021; Nagata et al., 2021), D-optimality is intended to minimize the volume of the confidence ellipsoid of state estimate (Saito et al., 2021; Yamada et al., 2021), and E-optimality is used to minimize the worst-case error (Nakai et al., 2021). Nakai et al. (2021) systematically compared these criteria under linear least-squares estimation and revealed that D-optimality achieves the lowest reconstruction error when the statistics of the system are well known, while A-optimality performs comparably to or better than D-optimality in more practical settings where this assumption does not hold; in both settings, E-optimality results in less accurate reconstruction than D-optimality and A-optimality. SSP has proven effective for state reconstruction; however, it does not directly target forecast error reduction.

In meteorological and geophysical sciences, the problem of designing observation networks to improve forecast accuracy in numerical weather prediction (NWP) has been studied widely in the ensemble-based data assimilation community based on the Ensemble Kalman Filter (EnKF; Evensen, 1994). The question of where to place supplementary observations to maximally reduce forecast error was first systematically explored by Lorenz and Emanuel (1998) using a low-dimensional chaotic model. Majumdar (2016) provides a comprehensive review of targeted observation campaigns. On the methodological side, Bishop and Toth (1999) proposed the Ensemble Transform (ET) technique to identify high-impact observation regions, and computationally efficient algorithms for this technique were subsequently developed by Zhang et al. (2016) and Wang et al. (2018). Bishop et al. (2001) and Majumdar et al. (2002) developed an Ensemble Transform Kalman Filter (ETKF)-based approach to determine observation locations that minimize the trace of the forecast error covariance matrix. Ancell and Hakim (2007) proposed an ensemble sensitivity analysis (ESA)-based approach to identify high-sensitivity observation locations by estimating forecast-error variance reduction. Yoshimura et al. (2020) introduced an observability Gramian from control theory to quantify the effectiveness of observation locations for state estimations. These approaches share a common focus on evaluating forecast-error metrics to determine observation locations. However, the framework of optimal experimental design based on the FIM has not been applied to observation network design in the context of data assimilation.

Although SSP and EnKF have been developed independently, they share a common mathematical structure. The ensemble perturbation subspace spanned by the background ensemble in EnKF is structurally equivalent to the singular vector subspace used in SSP, and the analysis error covariance matrix in EnKF can be interpreted as the inverse of the FIM, as elaborated in Sections 2 and 3. Despite these parallels, their connection has not been systematically formalized. It remains unclear how the FIMs in the ensemble space and model space differ when SSP is extended to the EnKF, and which is appropriate for designing observation networks that directly target forecast error reduction. While minimizing analysis errors is sufficient to minimize forecast errors in linear systems, this equivalence breaks down in nonlinear systems such as NWP models. This motivates designing observation networks that directly target forecast error reduction, rather than analysis error reduction alone. Furthermore, the mathematical interpretations of A-, D-, and E-optimality in terms of forecast error reduction within the EnKF context have not been clarified.

Beyond the formulation of the objective function, the observation network design problem is combinatorial and NP-hard in general. This makes brute-force search computationally infeasible, and greedy algorithms are therefore commonly employed in SSP as a practical alternative (e.g., Saito et al., 2021; Nakai et al., 2021; Yamada et al., 2021). Fast greedy algorithms have also been proposed in the SSP context (Saito et al., 2021; Nakai et al., 2021; Yamada et al., 2021), but their extension to the EnKF has not been explored.

The present study addresses these open questions by connecting SSP and EnKF through optimal experimental design. The main contributions of this study are threefold. First, while conventional SSP methods aim to reduce analysis errors, this study develops SSP for reducing forecast errors by using a tangent linear model approximated by an ensemble forecast. Second, we derive the FIMs in both the ensemble space and model space for the EnKF and clarify the mathematical interpretations of A-, D-, and E-optimality in terms of forecast error reduction. Third, we propose a new fast greedy algorithm for A-optimality in the model space (Fast AG-M), which avoids matrix inversion at each greedy step and substantially reduces computational cost relative to the baseline implementation. These results are validated through observing system simulation experiments (OSSEs) using the Lorenz-96 model.

The remainder of this paper is organized as follows. Section 2 reviews the theoretical background of SSP. Section 3 extends SSP to the EnKF. Section 4 introduces the greedy algorithms and proposes Fast AG-M. Section 5 presents OSSEs using the Lorenz-96 model. Section 6 summarizes the findings and discusses directions for future work. An overview of the paper structure is provided in Figure 1.

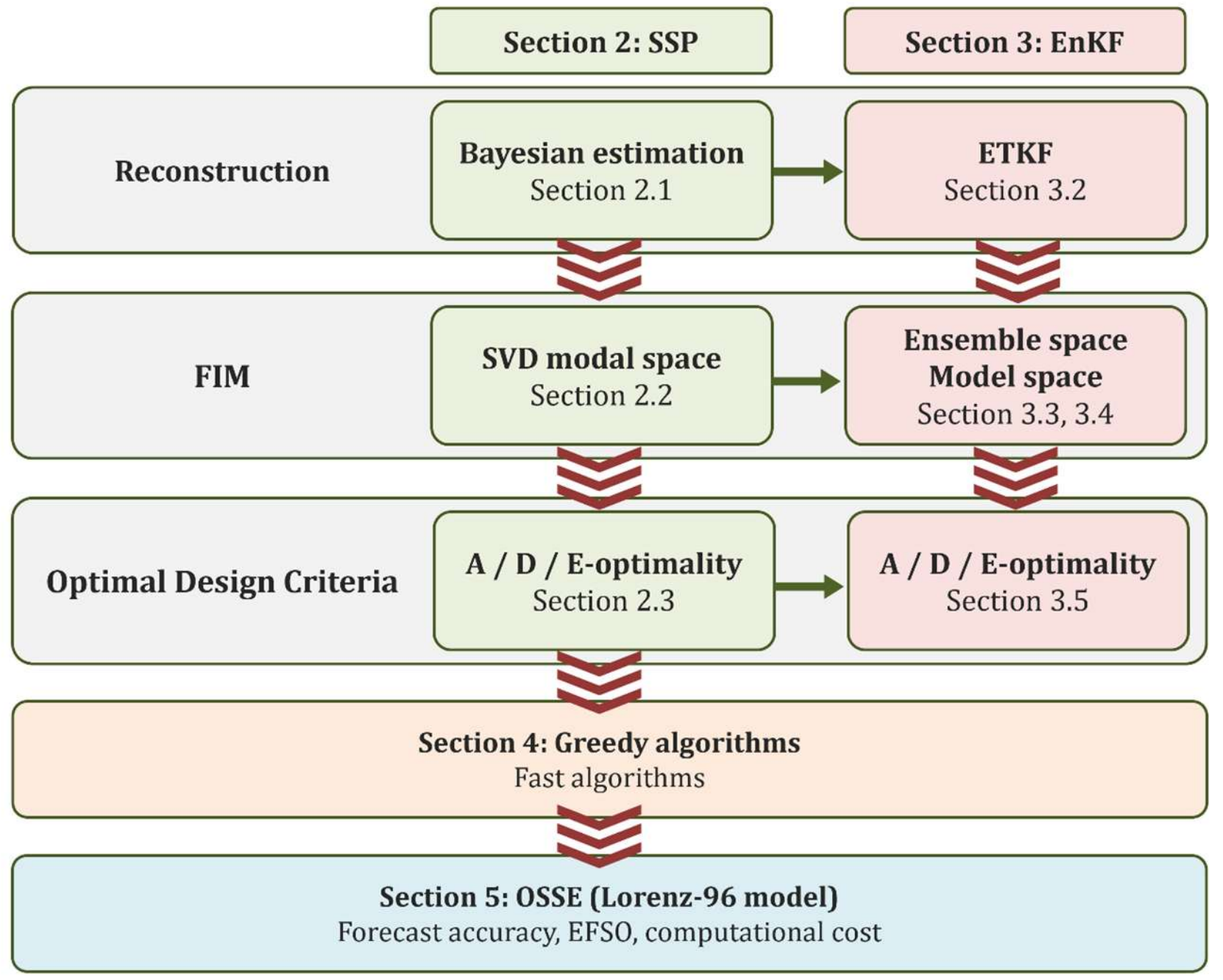

FIGURE 1 Overview of the paper structure. Horizontal arrows indicate the extension of the SSP to the EnKF. Vertical arrows indicate the logical flow within each framework.

## 2. Theoretical Background: Sparse Sensor Placement

This section reviews the theoretical background of SSP (Manohar et al., 2018), which determines observation locations by exploiting the low-dimensional structure in data via POD. Optimal sensor placement is formulated as the problem of maximizing a scalar measure of the FIM, which provides a lower bound on the reconstruction error through the Cramér–Rao inequality.

### 2.1 Reconstruction

In SSP, a system state is reconstructed from observations using a low-dimensional subspace obtained by POD of a training data matrix $\boldsymbol{X} = [\boldsymbol{x}_1 \quad \cdots \quad \boldsymbol{x}_{N_{\mathrm{train}}}] \in \mathbb{R}^{N_{\mathrm{model}} \times N_{\mathrm{train}}}$, which is performed via SVD. The rank-$r$ truncated SVD of $\boldsymbol{X} - \overline{\boldsymbol{x}}\boldsymbol{1}^\top$ yields

$$\boldsymbol{X} - \overline{\boldsymbol{x}}\boldsymbol{1}^\top \approx \widehat{\boldsymbol{U}}\widehat{\boldsymbol{\Sigma}}\widehat{\boldsymbol{V}}^\top, \tag{1}$$

where $\widehat{\boldsymbol{U}} \in \mathbb{R}^{N_{\mathrm{model}} \times r}$, $\widehat{\boldsymbol{\Sigma}} \in \mathbb{R}^{r \times r}$, $\widehat{\boldsymbol{V}} \in \mathbb{R}^{N_{\mathrm{train}} \times r}$, and $\overline{\boldsymbol{x}}$ is the sample mean. A state vector $\boldsymbol{x}$ is then approximated as

$$\boldsymbol{x} \approx \overline{\boldsymbol{x}} + \widehat{\boldsymbol{U}}\boldsymbol{z}, \tag{2}$$

where $\boldsymbol{z} \in \mathbb{R}^r$ is the vector of modal amplitudes. Given observations $\boldsymbol{y} = \boldsymbol{H}\boldsymbol{x} + \boldsymbol{\epsilon} \in \mathbb{R}^{N_{\mathrm{obs}}}$, where $\boldsymbol{H} \in \mathbb{R}^{N_{\mathrm{obs}} \times N_{\mathrm{model}}}$ is the linear observation operator and $\boldsymbol{\epsilon} \sim \mathcal{N}(\boldsymbol{0}, \boldsymbol{R})$ is Gaussian observation error with covariance $\boldsymbol{R}$, the reconstruction problem can be simplified to estimating $\boldsymbol{z}$ from $\boldsymbol{y}$ via the projected operator $\boldsymbol{C} = \boldsymbol{H}\widehat{\boldsymbol{U}}$. Following Yamada et al. (2021), the Bayesian estimate of $\boldsymbol{z}$ with prior covariance $\widehat{\boldsymbol{\Sigma}}^2/(N_{\mathrm{train}} - 1)$ is given by

$$\boldsymbol{z}_{\mathrm{BE}} = \left[(N_{\mathrm{train}} - 1)\widehat{\boldsymbol{\Sigma}}^{-2} + \boldsymbol{C}^\top \boldsymbol{R}^{-1} \boldsymbol{C}\right]^{-1} \boldsymbol{C}^\top (\boldsymbol{y} - \boldsymbol{H}\overline{\boldsymbol{x}}). \tag{3}$$

The corresponding state estimate is $\boldsymbol{x}_{\mathrm{BE}} = \overline{\boldsymbol{x}} + \widehat{\boldsymbol{U}}\boldsymbol{z}_{\mathrm{BE}}$. As shown in Section 3.2, this expression is equivalent to the ETKF analysis step when no dimensional truncation is applied.

### 2.2 Fisher Information Matrix

The FIM (Fisher, 1922) quantifies the expected amount of information that observations provide about unknown parameters, here taken to be the state vector $\boldsymbol{x}$. The inverse of the FIM provides the Cramér–Rao lower bound on the covariance of any unbiased estimator (Rodgers, 2000),

$$\mathbb{E}[(\boldsymbol{x} - \boldsymbol{x}_{\mathrm{true}})(\boldsymbol{x} - \boldsymbol{x}_{\mathrm{true}})^\top] \succeq \boldsymbol{\mathcal{F}}^{-1}, \tag{4}$$

where $\boldsymbol{\mathcal{F}}$ denotes the FIM and the matrix inequality $\boldsymbol{A} \succeq \boldsymbol{B}$ indicates that $\boldsymbol{A} - \boldsymbol{B}$ is positive semidefinite. When the posterior distribution is Gaussian, the Cramér–Rao bound is achieved with equality, and the FIM coincides with the inverse of the posterior error covariance matrix. The FIM with respect to the state vector $\boldsymbol{x}$ is defined as the negative expected Hessian of the log-posterior,

$$\boldsymbol{\mathcal{F}} = -\mathbb{E}\left[\left(\frac{\partial^2}{\partial \boldsymbol{x}^2}\log p(\boldsymbol{x}|\boldsymbol{y})\right)\right], \tag{5}$$

where the expectation is taken with respect to the posterior distribution $p(\boldsymbol{x}|\boldsymbol{y})$.

Applying this formulation to the reconstruction problem of Section 2.1, the FIM for the Bayesian estimator is given by the Hessian of the log-posterior (Yamada et al., 2021),

$$\boldsymbol{\mathcal{F}}_{\mathrm{BE}} = (N_{\mathrm{train}} - 1)\widehat{\boldsymbol{\Sigma}}^{-2} + \boldsymbol{C}^{\top}\boldsymbol{R}^{-1}\boldsymbol{C}. \tag{6}$$

Since the FIM quantifies the information that observations provide about unknown parameters, maximizing a scalar measure of $\boldsymbol{\mathcal{F}}_{\mathrm{BE}}$ serves as a natural objective for sensor placement. The sensor placement problem is therefore formulated as an optimal experimental design problem in which a set of observation locations is determined to maximize a scalar measure of the FIM given a constraint on the number of available sensors. As discussed in Section 2.3, however, there are several ways to define such a scalar measure of the FIM, each leading to a different optimality criterion.

### 2.3 Optimal Design Criteria for Sensor Placement

Representative optimal design criteria include A-, D-, and E-optimality (Pukelsheim, 2006). Let $\lambda_1, \lambda_2, \cdots, \lambda_r$ denote the eigenvalues of $\boldsymbol{\mathcal{F}}$, where $r$ is the rank of the FIM. We denote the objective function for each criterion as $\phi_{\mathrm{A}}$, $\phi_{\mathrm{D}}$, and $\phi_{\mathrm{E}}$, respectively. First, A-optimality seeks to minimize the trace of the inverse of the FIM,

$$\boldsymbol{H}_{\mathrm{A}} = \arg\min_{\boldsymbol{H}} \phi_{\mathrm{A}} = \arg\min_{\boldsymbol{H}} \mathrm{tr}\,\boldsymbol{\mathcal{F}}^{-1} = \arg\min_{\boldsymbol{H}} \sum_{i=1}^{r} \lambda_i^{-1}, \tag{7}$$

which corresponds to minimizing the average variance of the estimated state. Second, D-optimality minimizes the log-determinant of the inverse of the FIM,

$$\boldsymbol{H}_{\mathrm{D}} = \arg\min_{\boldsymbol{H}} \phi_{\mathrm{D}} = \arg\min_{\boldsymbol{H}} \log\det\boldsymbol{\mathcal{F}}^{-1} = \arg\min_{\boldsymbol{H}} \log\prod_{i=1}^{r} \lambda_i^{-1}, \tag{8}$$

which minimizes the volume of the confidence ellipsoid of the estimate. Finally, E-optimality minimizes the maximum eigenvalue of the inverse of the FIM,

$$\boldsymbol{H}_{\mathrm{E}} = \arg\min_{\boldsymbol{H}} \phi_{\mathrm{E}} = \arg\min_{\boldsymbol{H}} \lambda_{\max}(\boldsymbol{\mathcal{F}}^{-1}), \tag{9}$$

which targets the worst-case estimation accuracy by improving the most uncertain direction in the state space (Nakai et al., 2021).

In SSP studies, A-, D-, and E-optimality have been applied depending on the problem. In general, D-optimality is the most widely adopted criterion in the field of optimal experimental design (e.g., Saito et al., 2021; Yamada et al., 2021). A-optimality is preferred when minimizing the mean squared error is the primary concern (e.g., Nagata et al., 2021), while E-optimality minimizes the worst-case variance of estimation error (e.g., Nakai et al., 2021). As discussed later, we demonstrate that A-optimality is particularly appropriate for reducing forecast RMSE, which is a primary objective in NWP.

## 3. Sparse Sensor Placement for the Ensemble Kalman Filter

This section proposes new SSP methods for improving forecast accuracy by extending the optimal experimental design introduced in Section 2 to the EnKF. We first formulate the adaptive observation problem within the data assimilation cycle. We then demonstrate the equivalence between the Bayesian reconstruction in the SVD modal space and the ETKF analysis step, and derive the FIMs in both the ensemble space and the model space, showing that the model-space FIM is appropriate for directly targeting forecast error reduction. Finally, we present the A-, D-, and E-optimality criteria and clarify their mathematical interpretations in terms of forecast error reduction.

### 3.1 Observation Selection Problem in the Data Assimilation Cycle

This section formulates the problem of selecting observation locations within an ensemble-based data assimilation cycle. As illustrated in Figure 2, let the current time be $t = -t_c$, and consider the problem of determining the observation operator $\boldsymbol{H}_0$ at the assimilation time $t = 0$ so as to optimize a chosen measure of the forecast error covariance matrix $\boldsymbol{P}^f_{t_f}$ at the forecast time $t = t_f$.

At the current time, the analysis ensemble $\boldsymbol{X}^a_{-t_c} \in \mathbb{R}^{N_{\mathrm{model}} \times N_{\mathrm{ens}}}$ is assumed to be available. Propagating this ensemble forward to $t = 0$ yields the background ensemble $\boldsymbol{X}^b_0$, with mean $\overline{\boldsymbol{x}}^b_0$ and perturbations $\delta\boldsymbol{X}^b_0 = \boldsymbol{X}^b_0 - \overline{\boldsymbol{x}}^b_0 \boldsymbol{1}^\top$. The observation $\boldsymbol{y}_0$, the analysis ensemble $\boldsymbol{X}^a_0$, and the forecast ensemble $\boldsymbol{X}^f_{t_f}$ are all unavailable at the current time $t = -t_c$ since the assimilation time $t = 0$ lies in the future.

A key consideration in this setting concerns the choice of TLM. Minimizing the forecast error covariance matrix $\boldsymbol{P}^f_{t_f}$ would in principle require the TLM linearized around the trajectory after data assimilation, $\boldsymbol{M}^a$. However, $\boldsymbol{M}^a$ depends on the analysis state and is unavailable at the current time. We instead employ the surrogate TLM $\boldsymbol{M}^b$ linearized around the background trajectory. The forecast ensemble perturbations $\delta\boldsymbol{X}^f_{t_f}$ are consequently approximated by $\delta\boldsymbol{X}_{t_f}^b \approx \boldsymbol{M}^b \delta\boldsymbol{X}_0^b$, obtained by propagating the background ensemble $\boldsymbol{X}_0^b$ to the forecast time $t = t_f$. This surrogate approach allows the same $\boldsymbol{M}^b$ to be used across all candidate observation locations, since $\boldsymbol{M}^b$ is independent of the chosen locations.

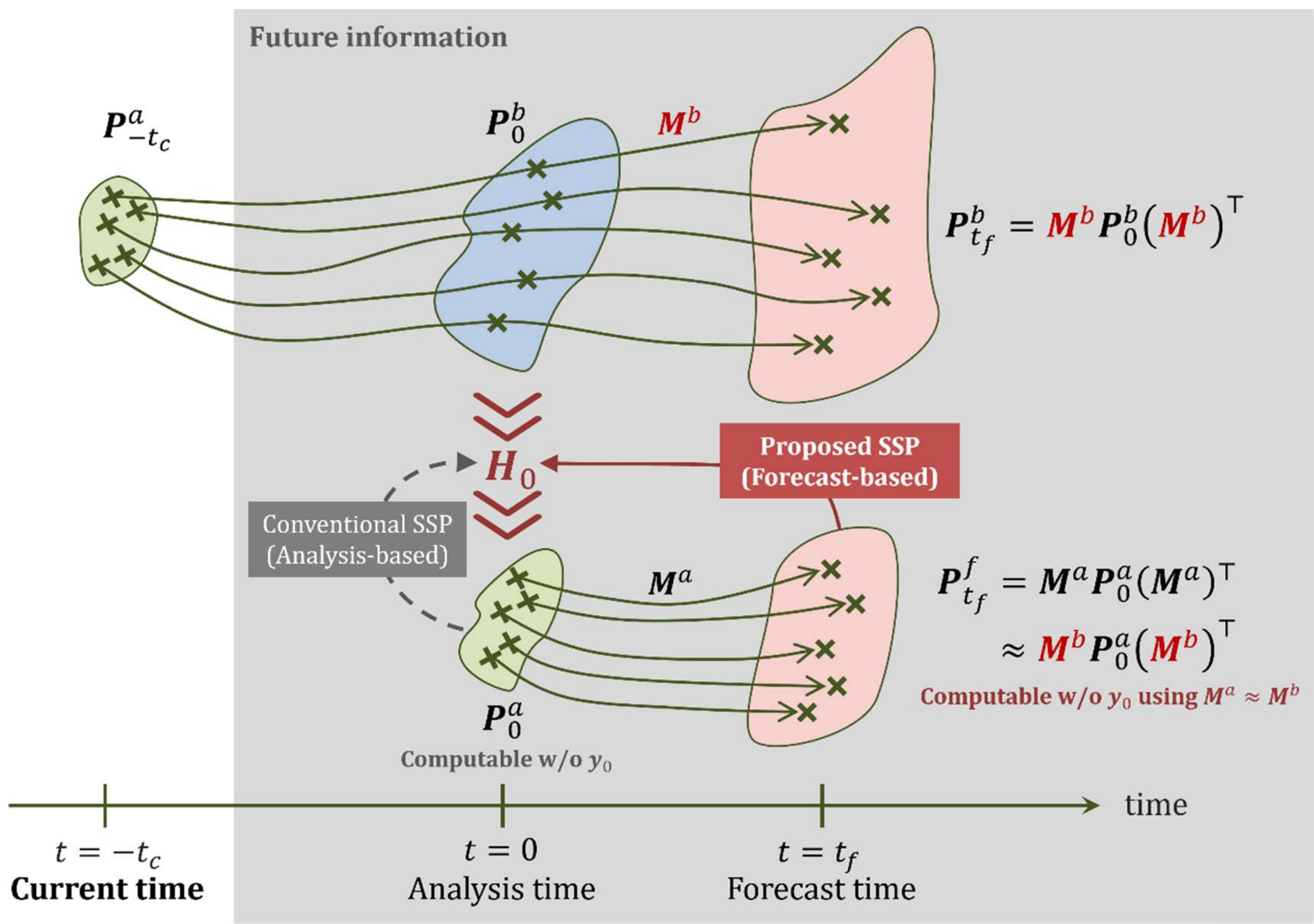

FIGURE 2 Schematic overview of the adaptive observation problem in an ensemble-based data assimilation cycle. The gray region indicates future information relative to the current time $t = -t_c$. The analysis error covariance matrix $\boldsymbol{P}_{-t_c}^a$ at the current time propagates to the background error covariance matrix $\boldsymbol{P}_0^b$ at the analysis time $t = 0$. Since the analysis error covariance matrix $\boldsymbol{P}_0^a$ can be computed without observation $\boldsymbol{y}_0$, conventional SSP methods determine the observation locations $\boldsymbol{H}_0$ that minimizes $\boldsymbol{P}_0^a$. In the proposed approach, $\boldsymbol{P}_0^a$ is further propagated to the forecast time $t = t_f$ to evaluate the forecast error covariance $\boldsymbol{P}_{t_f}^f$ using the surrogate TLM $\boldsymbol{M}^b$ to determine the observation locations $\boldsymbol{H}_0$ that minimizes forecast errors at $t = t_f$.

### 3.2 Reconstruction by Ensemble Transform Kalman Filter

Data assimilation can be regarded as a reconstruction procedure in the context of SSP. In the ETKF (Bishop et al., 2001), the background ensemble perturbations $\delta\boldsymbol{X}_0^b$ are interpreted as a linear mapping from the ensemble space $\mathbb{R}^{N_{\text{ens}}}$ to the model space $\mathbb{R}^{N_{\text{model}}}$ (Hunt et al., 2007). Assuming that the ensemble-space state vector $\boldsymbol{w}_0 \in \mathbb{R}^{N_{\text{ens}}}$ follows a Gaussian prior with covariance $\widetilde{\boldsymbol{P}}_0^b = \boldsymbol{I}/(N_{\text{ens}} - 1)$, the posterior distribution given observations $\boldsymbol{y}_0$ yields the analysis in the ensemble space,

$$\boldsymbol{w}_0^a = \left[(N_{\text{ens}} - 1)\boldsymbol{I} + \left(\boldsymbol{H}_0\delta\boldsymbol{X}_0^b\right)^\top \boldsymbol{R}_0^{-1}\left(\boldsymbol{H}_0\delta\boldsymbol{X}_0^b\right)\right]^{-1}\left(\boldsymbol{H}_0\delta\boldsymbol{X}_0^b\right)^\top\left(\boldsymbol{y}_0 - \boldsymbol{H}_0\overline{\boldsymbol{x}}_0^b\right), \tag{10}$$

and the analysis ensemble mean is given by $\overline{\boldsymbol{x}}_0^a = \overline{\boldsymbol{x}}_0^b + \delta\boldsymbol{X}_0^b\boldsymbol{w}_0^a$.

The Bayesian estimation in the SVD modal space introduced in Section 2 and the ETKF perform essentially identical computations. In the model-space formulation, the standard Kalman filter equations give

$$(\boldsymbol{P}_0^a)^{-1} = \left(\boldsymbol{P}_0^b\right)^{-1} + \boldsymbol{H}_0^\top \boldsymbol{R}_0^{-1}\boldsymbol{H}_0. \tag{11}$$

The ensemble-space and SVD modal-space formulations satisfy analogous equations, with $\delta\boldsymbol{X}_0^b$ and $\widehat{\boldsymbol{U}}_0$ as the respective basis matrices. The resulting analysis error covariance matrices in the ensemble space and SVD modal space are given by

$$\left(\widetilde{\boldsymbol{P}}_0^a\right)^{-1} = (N_{\text{ens}} - 1)\boldsymbol{I} + \left(\boldsymbol{H}_0\delta\boldsymbol{X}_0^b\right)^\top \boldsymbol{R}_0^{-1}\left(\boldsymbol{H}_0\delta\boldsymbol{X}_0^b\right) \in \mathbb{R}^{N_{\text{ens}}\times N_{\text{ens}}}, \tag{12}$$

$$\left(\widehat{\boldsymbol{P}}_0^a\right)^{-1} = (N_{\text{ens}} - 1)\widehat{\boldsymbol{\Sigma}}^{-2} + \left(\boldsymbol{H}_0\widehat{\boldsymbol{U}}_0\right)^\top \boldsymbol{R}_0^{-1}\left(\boldsymbol{H}_0\widehat{\boldsymbol{U}}_0\right) \in \mathbb{R}^{r\times r}. \tag{13}$$

When no dimensional truncation is applied, these three formulations coincide:

$$\boldsymbol{P}_0^a = \delta\boldsymbol{X}_0^b\widetilde{\boldsymbol{P}}_0^a\left(\delta\boldsymbol{X}_0^b\right)^\top = \boldsymbol{U}_0\widehat{\boldsymbol{P}}_0^a\boldsymbol{U}_0^\top. \tag{14}$$

These relationships are summarized in Figure 3. From the above discussion, it follows that the SSP developed in Section 2 for low-dimensional modal spaces can also be applied to the ETKF formulation in ensemble space.

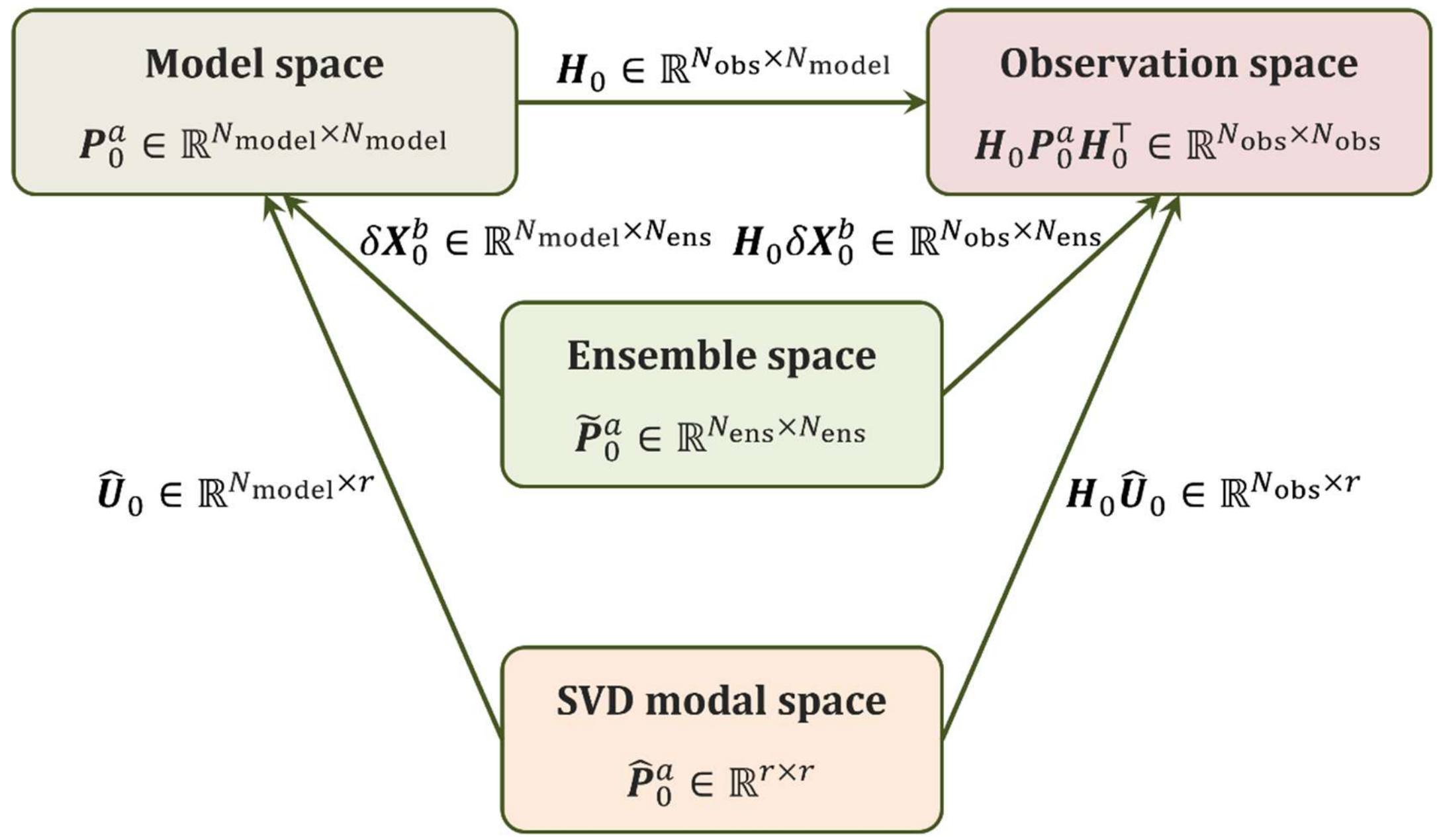

FIGURE 3 Relationships of analysis error covariance in the model space, ensemble space, SVD modal space, and observation space, together with the corresponding mapping matrices.

The analysis error covariance matrices in each space are shown inside the boxes. This figure extends the space relationships summarized in Kotsuki et al. (2020) by additionally incorporating the SVD modal space introduced in the SSP.

### 3.3 Fisher Information Matrix in the Ensemble Space

Here we derive the FIM in the ensemble space. Assuming a Gaussian posterior distribution, the FIM is given by the negative Hessian of the log-posterior, analogously to Eq. (5). Under the ETKF formulation, the FIM for the analysis in the ensemble space is given by the inverse of the analysis error covariance matrix,

$$\widetilde{\boldsymbol{\mathcal{F}}}_0^a = \left(\widetilde{\boldsymbol{P}}_0^a\right)^{-1} = (N_{\text{ens}} - 1)\boldsymbol{I} + \left(\boldsymbol{H}_0 \delta \boldsymbol{X}_0^b\right)^{\top} \boldsymbol{R}_0^{-1} \left(\boldsymbol{H}_0 \delta \boldsymbol{X}_0^b\right). \tag{15}$$

However, the FIM in the ensemble space is defined only at the analysis time, because the time evolution of the state cannot be represented in the ensemble space under the EnKF. The ensemble-space FIM is therefore useful for reducing analysis error, but does not directly characterize forecast error reduction. To design observation networks that directly target forecast error reduction, it is necessary to define a FIM at the forecast time, which requires working in the model space as described in Section 3.4.

### 3.4 Fisher Information Matrix in the Model Space

We derive the FIM in the model space to directly characterize forecast error reduction. Analogously to Section 3.3, the FIM associated with the analysis state is given by the inverse of the analysis error covariance matrix (Rodgers, 2000; Bocquet, 2009),

$$\boldsymbol{\mathcal{F}}_0^a = (\boldsymbol{P}_0^a)^{-1} = \left(\boldsymbol{P}_0^b\right)^{-1} + \boldsymbol{H}_0^{\top} \boldsymbol{R}_0^{-1} \boldsymbol{H}_0. \tag{16}$$

Similarly, the FIM associated with the forecast state at $t = t_f$ is given by the inverse of the forecast error covariance matrix,

$$\boldsymbol{\mathcal{F}}_{t_f}^f = \left(\boldsymbol{P}_{t_f}^f\right)^{-1} \approx (\boldsymbol{M}^a \boldsymbol{P}_0^a (\boldsymbol{M}^a)^{\top})^{-1}, \tag{17}$$

where $\boldsymbol{M}^a$ denotes the TLM linearized around the analysis trajectory. As discussed in Section 3.1, however, $\boldsymbol{M}^a$ is unavailable in practice and is replaced by $\boldsymbol{M}^b$. Using the ensemble approximation $\delta \boldsymbol{X}_{t_f}^b \approx \boldsymbol{M}^b \delta \boldsymbol{X}_0^b$, the inverse of the FIM for the forecast state is expressed as

$$\left(\boldsymbol{\mathcal{F}}_{t_f}^f\right)^{-1} = \boldsymbol{P}_{t_f}^f \approx \delta \boldsymbol{X}_{t_f}^b \widetilde{\boldsymbol{P}}_0^a \left(\delta \boldsymbol{X}_{t_f}^b\right)^{\top}. \tag{18}$$

Consequently, designing observation networks that directly target forecast error reduction requires the model-space FIM, Eq. (18), rather than the ensemble-space FIM, Eq. (15).

### 3.5 Optimal Design Criteria for the Ensemble Kalman Filter

In this section, we introduce objective functions for observation network design aimed at improving forecast accuracy, based on the FIMs in the ensemble space and the model space derived in Sections 3.3 and 3.4. As shown in Section 3.4, the model-space FIM is the appropriate choice for directly targeting forecast error reduction. We therefore present A- and E-optimality in the model space, and discuss why D-optimality must be formulated in the ensemble space rather than model space. The properties of each optimality criterion are summarized in Table 1.

TABLE 1 Properties of A-, D-, and E-optimality for the EnKF.

| Optimality | Space | Property |
|---|---|---|
| A-optimality | Model | Minimizes the mean forecast error variance |
| D-optimality | Model | Ill-defined (see Section 3.5.2) |
| | Ensemble | Maximizes the Shannon information content of the assimilated observations |
| E-optimality | Model | Minimizes the worst-case variance of forecast error |

### 3.5.1 A-optimality

The objective function based on A-optimality in the model space is the trace of the forecast error covariance matrix, defined as

$$\boldsymbol{H}_{0,\mathrm{A|Model}} = \arg\min_{\boldsymbol{H}_0} \phi_{\mathrm{A|Model}} = \arg\min_{\boldsymbol{H}_0} \mathrm{tr}\left(\boldsymbol{P}_{t_f}^{f}\right) = \arg\min_{\boldsymbol{H}_0} \mathrm{tr}\left(\delta\boldsymbol{X}_{t_f}^{b}\widetilde{\boldsymbol{P}}_0^{a}\left(\delta\boldsymbol{X}_{t_f}^{b}\right)^{\top}\right). \quad (19)$$

This criterion minimizes the mean forecast error variance across all state components, which is equivalent to minimization of the sum of eigenvalues of the forecast error covariance matrix. This coincides with the method proposed by Bishop et al. (2001).

### 3.5.2 D-optimality

The objective function based on D-optimality in the model space is the log-determinant of the forecast error covariance matrix, defined as

$$\boldsymbol{H}_{0,\mathrm{D|Model}} = \arg\min_{\boldsymbol{H}_0} \phi_{\mathrm{D|Model}} = \arg\min_{\boldsymbol{H}_0} \log\det \boldsymbol{P}_{t_f}^{f}. \quad (20)$$

However, D-optimality in the model space is ill-defined for two reasons. First, since the surrogate TLM $\boldsymbol{M}^b$ is independent of the choice of observation locations, the multiplicative property of the determinant gives

$$\boldsymbol{H}_{0,\mathrm{D|Model}} = \arg\min_{\boldsymbol{H}_0} \phi_{\mathrm{D|Model}} = \arg\min_{\boldsymbol{H}_0} \log\det \boldsymbol{P}_{t_f}^{f} = \arg\min_{\boldsymbol{H}_0} \log\det \boldsymbol{P}_0^{a}, \quad (21)$$

because $\det \boldsymbol{P}_{t_f}^{f} = \det(\boldsymbol{M}^b \boldsymbol{P}_0^a (\boldsymbol{M}^b)^\top) = \det \boldsymbol{M}^b \det \boldsymbol{P}_0^a \det(\boldsymbol{M}^b)^\top$ by the multiplicative

property of the determinant. This optimization therefore reduces to minimizing the determinant of the analysis error covariance matrix rather than the forecast error covariance matrix. Second, under the EnKF, the analysis error covariance matrix satisfies $\mathrm{rk}\,\boldsymbol{P}_0^a \leq N_{\mathrm{ens}} - 1$, meaning that $\boldsymbol{P}_0^a$ is rank-deficient in typical applications where $N_{\mathrm{model}} \gg N_{\mathrm{ens}}$. Consequently, $\det \boldsymbol{P}_0^a = 0$ for any choice of observation locations, making the objective function ill-defined.

Consequently, D-optimality is instead formulated in the ensemble space,

$$\boldsymbol{H}_{0,\mathrm{D|Ens}} = \arg\min_{\boldsymbol{H}_0} \phi_{\mathrm{D|Ens}} = \arg\min_{\boldsymbol{H}_0} \log\det \widetilde{\boldsymbol{P}}_0^a = \arg\max_{\boldsymbol{H}_0} \log\det\left(\widetilde{\boldsymbol{P}}_0^a\right)^{-1}, \tag{22}$$

which is guaranteed to be well-defined under the ETKF formulation since $\widetilde{\boldsymbol{P}}_0^a$ is positive definite. This criterion is equivalent to maximizing the Shannon information content of the assimilated observations in the ensemble space (Rodgers, 2000, Fowler and van Leeuwen, 2012, 2013).

### 3.5.3 E-optimality

The objective function based on E-optimality in the model space is the largest eigenvalue of the forecast error covariance matrix, defined as

$$\boldsymbol{H}_{0,\mathrm{E|Model}} = \arg\min_{\boldsymbol{H}_0} \phi_{\mathrm{E|Model}} = \arg\min_{\boldsymbol{H}_0} \lambda_{\max}\left(\boldsymbol{P}_{t_f}^f\right) = \arg\min_{\boldsymbol{H}_0} \lambda_{\max}\left(\delta\boldsymbol{X}_{t_f}^b \widetilde{\boldsymbol{P}}_0^a \left(\delta\boldsymbol{X}_{t_f}^b\right)^{\top}\right). \tag{23}$$

This criterion minimizes the worst-case variance of forecast error, thereby targeting the dominant direction of forecast error growth (Nakai et al., 2021).

## 4. Greedy Algorithms

This section proposes and examines a fast greedy algorithm for observation network design. Greedy algorithms, which iteratively select the locally optimal choice at each step, yield suboptimal solutions in general but are widely used in combinatorial optimization problems due to their computational efficiency and empirical effectiveness. In this study, we refer to the greedy methods for A-, D-, and E-optimality as AG, DG, and EG, respectively, appending "-M" when the FIM is defined in the model space and "-E" when defined in the ensemble space. For example, AG-M denotes the greedy method for A-optimality in the model space. Throughout this section, we assume that each measurement corresponds to a single scalar observation at a model grid point, so that the observation operator is binary, with elements taking values of either 0 or 1.

### 4.1 Greedy Selection Framework

Let $S_{\text{cand}}$ denote the index set of candidate locations for adaptive observations, and let $N_{\text{cand}} = |S_{\text{cand}}|$ denotes the number of candidate locations. For example, if $i$th, $j$th, and $k$th grid points are candidate locations, then $S_{\text{cand}} = \{i, j, k\}$ and $N_{\text{cand}} = 3$. The greedy algorithm sequentially selects $N_{\text{adp}}$ adaptive observation locations from $S_{\text{cand}}$ so as to optimize a chosen objective function, where $N_{\text{adp}}$ is the number of adaptive observations. In addition, let $S_{\text{fixed}}$ denote the index set of fixed non-adaptive observation locations.

At each greedy step $k\left(1 \leq k \leq N_{\text{adp}}\right)$, given the set of already selected locations $S_{k-1}$, the next observation location $i_k$ is chosen as

$$i_k = \arg \min_{i \in S_{\text{cand}} \setminus S_{k-1}} \phi(S_{k-1} \cup \{i\}). \tag{24}$$

Here, $S_{\text{cand}} \setminus S_{k-1}$ denotes the set of $k$th-step candidate locations that have not yet been selected up to step $k-1$, i.e., the remaining candidates available for $k$th-step selection. The function $\phi(S_{k-1} \cup \{i\})$ represents the value of the objective function when candidate location $i$ is tentatively added to the current selection $S_{k-1}$. The greedy algorithm then selects the location $i$ that minimizes this value over all remaining candidates $S_{\text{cand}} \setminus S_{k-1}$. Note that objective functions are defined to be minimized for A-, D-, and E-optimality (Eqs.(19), (22), (23)). After selecting $i_k$, the observation-location set is updated as $S_k = S_{k-1} \cup \{i_k\}$. Here, $S_0 = \emptyset$ holds before the first greedy step.

In this paper, the notation $A_k^{(i)}$ denotes the value of a quantity $A$ when the $i$th observation index is selected at the $k$th step, and $A_k = A_k^{(i)}\big|_{i=i_k}$ denotes its value after the $k$th selection has been determined. The flowchart of the AG-M algorithm is summarized in Algorithm 1. The other methods, DG-E and EG-M, can be implemented using the same algorithmic framework by replacing $\phi_{\text{A|Model}}$ with $\phi_{\text{D|Ens}}$ and $\phi_{\text{E|Model}}$, respectively.

---

**Algorithm 1.** AG-M algorithm

---

Input: $S_{\text{fixed}}, S_{\text{cand}}, \delta\boldsymbol{X}_0^b, \delta\boldsymbol{X}_{t_f}^b$

Output: $S_{N_{\text{adp}}}$

$\boldsymbol{H}_{0,0}, \boldsymbol{R}_{0,0} \leftarrow$ observation operator and observation error covariance matrix for $S_{\text{fixed}}$

$S_0 \leftarrow \emptyset$

For $k = 1, \cdots, N_{\text{adp}}$:

$$i_k \leftarrow \arg \min_{i \in S_{\text{cand}} \setminus S_{k-1}} \phi_{\text{A|Model}}$$

$$= \arg \min_{i \in S_{\text{cand}} \setminus S_{k-1}} \operatorname{tr}\left(\delta\boldsymbol{X}_{t_f}^b \left[(N_{\text{ens}} - 1)\boldsymbol{I} + \left(\boldsymbol{H}_{0,k}^{(i)}\, \delta\boldsymbol{X}_0^b\right)^\top \left(\boldsymbol{R}_{0,k}^{(i)}\right)^{-1} \left(\boldsymbol{H}_{0,k}^{(i)}\, \delta\boldsymbol{X}_0^b\right)\right]^{-1} \left(\delta\boldsymbol{X}_{t_f}^b\right)^\top\right)$$

$S_k \leftarrow S_{k-1} \cup \{i_k\}$

end

---

### 4.2 Fast Algorithm

In this section, we develop a fast algorithm for AG-M and interpret the fast algorithm of Yamada et al. (2021) for DG-E in the context of data assimilation. We introduce the following notation. The background ensemble perturbations $\delta\boldsymbol{X}_0^b$ are represented by the row vectors $\boldsymbol{w}^{(i)\top} \in \mathbb{R}^{N_{\mathrm{ens}}}$ as

$$\delta\boldsymbol{X}_0^b = \begin{bmatrix} \boldsymbol{w}^{(1)\top} \\ \vdots \\ \boldsymbol{w}^{(N_{\mathrm{model}})\top} \end{bmatrix}. \tag{25}$$

At the $k$th greedy step, the projected background ensemble perturbations are denoted by $\boldsymbol{C}_k^{(i)} = \boldsymbol{H}_k^{(i)}\delta\boldsymbol{X}_0^b$. The observation error covariance matrix $\boldsymbol{R}_{0,k}^{(i)}$ at step $k$ is expressed in block form as

$$\boldsymbol{R}_{0,k}^{(i)} = \begin{bmatrix} \boldsymbol{R}_{0,k-1} & \boldsymbol{r}_k^{(i)} \\ \left(\boldsymbol{r}_k^{(i)}\right)^\top & \left(\sigma_k^{(i)}\right)^2 \end{bmatrix}, \tag{26}$$

where $\boldsymbol{r}_k^{(i)} \in \mathbb{R}^{(N_0+k-1)}$ is the cross-covariance vector and $\left(\sigma_k^{(i)}\right)^2$ is the observation error variance at the $i$th candidate location. Using this block structure, we define the Schur complement $s_k^{(i)} \in \mathbb{R}$ and the vector $\boldsymbol{d}_k^{(i)} \in \mathbb{R}^{N_{\mathrm{ens}}}$ as

$$s_k^{(i)} = \left(\sigma_k^{(i)}\right)^2 - \left(\boldsymbol{r}_k^{(i)}\right)^\top \boldsymbol{R}_{0,k-1}^{-1}\boldsymbol{r}_k^{(i)}, \tag{27}$$

$$\boldsymbol{d}_k^{(i)} = \frac{1}{\sqrt{s_k^{(i)}}}\left[\boldsymbol{C}_{k-1}^\top \boldsymbol{R}_{0,k-1}^{-1}\boldsymbol{r}_k^{(i)} - \boldsymbol{w}^{(i)}\right]. \tag{28}$$

#### 4.2.1 Fast AG-M algorithm

In this study, we propose a fast algorithm for AG-M that is mathematically equivalent to the baseline AG-M but avoids matrix inversion at each greedy step. By reformulating the objective function using the Sherman–Morrison–Woodbury formula (see APPENDIX for details), the matrix inversion of $\widetilde{\boldsymbol{P}}_{0,k-1}^a$ at each greedy step can be replaced by an inexpensive scalar reciprocal operation. The resulting fast algorithm selects the $k$th observation location as

$$i_k = \arg\max_{i\in S_{\mathrm{cand}}\setminus S_{k-1}} \mathrm{tr}\left[\frac{\left(\widetilde{\boldsymbol{P}}_{0,k-1}^a \boldsymbol{d}_k^{(i)}\right)\left(\widetilde{\boldsymbol{P}}_{0,k-1}^a \boldsymbol{d}_k^{(i)}\right)^\top}{1+\left(\boldsymbol{d}_k^{(i)}\right)^\top\left(\widetilde{\boldsymbol{P}}_{0,k-1}^a \boldsymbol{d}_k^{(i)}\right)}\left(\delta\boldsymbol{X}_{t_f}^b\right)^\top \delta\boldsymbol{X}_{t_f}^b\right]. \tag{29}$$

When the observation error covariance matrix is diagonal, $\boldsymbol{d}_k^{(i)}$ reduces to $\boldsymbol{d}_k^{(i)} = -\boldsymbol{w}^{(i)}/\sigma_k^{(i)}$, and Eq. (29) simplifies to

$$i_k = \arg\max_{i\in S_{\mathrm{cand}}\setminus S_{k-1}} \mathrm{tr}\left[\frac{\left(\widetilde{\boldsymbol{P}}^a_{0,k-1}\boldsymbol{w}^{(i)}\right)\left(\widetilde{\boldsymbol{P}}^a_{0,k-1}\boldsymbol{w}^{(i)}\right)^\top}{\left(\sigma_k^{(i)}\right)^2 + \boldsymbol{w}^{(i)\top}\left(\widetilde{\boldsymbol{P}}^a_{0,k-1}\boldsymbol{w}^{(i)}\right)}\left(\delta\boldsymbol{X}^b_{t_f}\right)^\top \delta\boldsymbol{X}^b_{t_f}\right]. \tag{30}$$

The flowchart of the fast AG-M algorithm is summarized in Algorithm 2. Since Fast AG-M is mathematically equivalent to the baseline AG-M, it yields identical observation location selections with reduced computational cost.

---

**Algorithm 2.** Fast AG-M algorithm

---

Input: $S_{\mathrm{fixed}}, S_{\mathrm{cand}}, \delta\boldsymbol{X}^b_0, \delta\boldsymbol{X}^b_{t_f}$

Output: $S_{N_{\mathrm{adp}}}$

$\boldsymbol{H}_{0,0}, \boldsymbol{R}_{0,0} \leftarrow$ observation operator and observation error covariance matrix for $S_{\mathrm{fixed}}$

$S_0 \leftarrow \emptyset$

Calculate $\left(\delta\boldsymbol{X}^b_{t_f}\right)^\top \delta\boldsymbol{X}^b_{t_f}$

For $k = 1, \cdots, N_{\mathrm{adp}}$:

$$i_k \leftarrow \arg\max_{i\in S_{\mathrm{cand}}\setminus S_{k-1}} \mathrm{tr}\left[\frac{\left(\widetilde{\boldsymbol{P}}^a_{0,k-1}\boldsymbol{d}_k^{(i)}\right)\left(\widetilde{\boldsymbol{P}}^a_{0,k-1}\boldsymbol{d}_k^{(i)}\right)^\top}{1 + \left(\boldsymbol{d}_k^{(i)}\right)^\top\left(\widetilde{\boldsymbol{P}}^a_{0,k-1}\boldsymbol{d}_k^{(i)}\right)}\left(\delta\boldsymbol{X}^b_{t_f}\right)^\top \delta\boldsymbol{X}^b_{t_f}\right]$$

$S_k \leftarrow S_{k-1} \cup \{i_k\}$

$\widetilde{\boldsymbol{P}}^a_{0,k} \leftarrow \left[(N_{\mathrm{ens}} - 1)\boldsymbol{I} + \boldsymbol{C}_k^\top \boldsymbol{R}_{0,k}^{-1}\boldsymbol{C}_k\right]^{-1}$

end

---

### 4.2.2 The fast DG-E algorithm

Applying the fast algorithm of Yamada et al. (2021) to the ensemble space yields Fast DG-E, which is mathematically equivalent to the baseline DG-E. The $k$th observation location for DG-E is selected as

$$i_k = \arg\max_{i\in S_{\mathrm{cand}}\setminus S_{k-1}} \det\left(\widetilde{\boldsymbol{P}}^{a(i)}_{0,k}\right)^{-1} = \arg\max_{i\in S_{\mathrm{cand}}\setminus S_{k-1}} \frac{\left(\boldsymbol{d}_k^{(i)}\right)^\top \widetilde{\boldsymbol{P}}^a_{0,k-1}\boldsymbol{d}_k^{(i)}}{s_k^{(i)}}. \tag{31}$$

When the observation error covariance matrix is diagonal, Eq. (31) simplifies to

$$i_k = \arg\max_{i\in S_{\mathrm{cand}}\setminus S_{k-1}} \det\left(\widetilde{\boldsymbol{P}}^{a(i)}_{0,k}\right)^{-1} = \arg\max_{i\in S_{\mathrm{cand}}\setminus S_{k-1}} \frac{\boldsymbol{w}^{(i)\top}\widetilde{\boldsymbol{P}}^a_{0,k-1}\boldsymbol{w}^{(i)}}{\left(\sigma_k^{(i)}\right)^2}. \tag{32}$$

Since $\boldsymbol{w}^{(i)\top}\widetilde{\boldsymbol{P}}^a_{0,k-1}\boldsymbol{w}^{(i)}$ corresponds to the $i$th diagonal entry of the analysis error covariance matrix in the model space $\boldsymbol{P}^a_{0,k-1} = \delta\boldsymbol{X}^b_0\widetilde{\boldsymbol{P}}^a_{0,k-1}\left(\delta\boldsymbol{X}^b_0\right)^\top$, it represents the analysis error variance at the $i$th grid point. It therefore follows from Eq. (32) that DG-E selects at each greedy step the observation location at which the analysis error variance, normalized by the observation error variance $\left(\sigma_k^{(i)}\right)^2$, is largest given the observations chosen up to step $k-1$.

### 4.2.3 Computational Complexity of the Greedy Algorithms

Table 2 summarizes the computational complexity of the greedy algorithms. Throughout, we assume $N_{\mathrm{model}} > N_{\mathrm{ens}}$. As shown in Table 2, Fast AG-M reduces the computational complexity of the baseline AG-M by replacing the matrix inversion at each greedy step with an inexpensive scalar reciprocal operation, as described in Section 4.2.1. The fast DG-E algorithm also achieves low computational complexity by confining the dominant operations to ensemble-space multiplications. Since $N_{\mathrm{model}} \gg N_{\mathrm{ens}}$ in typical data assimilation problems, both methods scale efficiently to high-dimensional problems.

By contrast, EG-M incurs the highest computational cost among the methods compared, as it requires an eigenvalue decomposition of the full model-space covariance matrix at each greedy step. Consequently, the computational cost grows substantially as $N_{\mathrm{model}}$ increases, as summarized in Table 2.

TABLE 2 Computational complexity of the greedy algorithms.

| Method | Computational complexity |
|---|---|
| Brute-force | $\frac{N_{\mathrm{cand}}!}{(N_{\mathrm{cand}}-N_{\mathrm{adp}})!\,N_{\mathrm{adp}}!} \sim O((N_{\mathrm{cand}})^{N_{\mathrm{adp}}})$ |
| AG-M | $O(N_{\mathrm{adp}} \times N_{\mathrm{cand}} \times N_{\mathrm{model}}^2 \times N_{\mathrm{ens}})$ |
| Fast AG-M | $O(N_{\mathrm{adp}} \times N_{\mathrm{cand}} \times N_{\mathrm{ens}}^3)$ |
| DG-E | $O(N_{\mathrm{adp}} \times N_{\mathrm{cand}} \times N_{\mathrm{ens}}^3)$ |
| Fast DG-E | $O(N_{\mathrm{adp}} \times N_{\mathrm{cand}} \times N_{\mathrm{ens}}^2)$ |
| EG-M | $O(N_{\mathrm{adp}} \times N_{\mathrm{cand}} \times N_{\mathrm{model}}^3)$ |

## 5. Experiments

### 5.1 Experimental Setting

We evaluate the performance of the observation network design methods developed in the preceding sections through numerical experiments. Here, we conduct a series of OSSEs using the Lorenz-96 model (Lorenz, 1996),

$$\frac{dx_j}{dt} = (x_{j+1} - x_{j-2})x_{j-1} - x_j + F, \qquad j = 1, \cdots, 40, \tag{33}$$

with $N_{\text{model}} = 40$ grid points, forcing parameter $F = 8.0$, and time step $\Delta t = 0.05$. Following Lorenz and Emanuel (1998), four time steps correspond to one model day, and time integration is performed using a fourth-order Runge–Kutta scheme.

As shown in Figure 4, the 40 grid points are divided equally into land and ocean regions, with fixed observation sites over land and candidate adaptive observation sites over the ocean following the experimental setting of Lorenz and Emanuel (1998). Observations are generated by applying the observation operator $\boldsymbol{H}_0$ to the nature run and adding zero-mean Gaussian noise with unit variance.

Data assimilation is performed using the ETKF (Bishop et al., 2001) with ensemble size $N_{\text{ens}} = 32$, observation error covariance $\boldsymbol{R}_0 = \boldsymbol{I}$, and multiplicative inflation factor $\alpha = 1.01$. No localization is applied. The assimilation interval and forecast lead time are fixed at $t_c = 6$ hours and $t_f = 2$ days, respectively.

The experimental procedure consists of the following steps:

1. Prepare the analysis ensemble $\boldsymbol{X}^a_{-t_c}$ at the current time $t = -t_c$, obtained from an assimilation–forecast cycle using five randomly selected ocean observation sites together with the fixed twenty land observation sites.
2. Propagate the ensemble forward to obtain the background ensemble $\boldsymbol{X}^b_0$ at the assimilation time $t = 0$ and $\boldsymbol{X}^b_{t_f}$ at the forecast time $t = t_f$.
3. Select adaptive observation sites over the ocean using each greedy method (AG-M, Fast AG-M, DG-E, Fast DG-E, and EG-M), based on $\boldsymbol{X}^b_0$ and $\boldsymbol{X}^b_{t_f}$, and define the observation operator $\boldsymbol{H}_0$.
4. Generate the observations $\boldsymbol{y}_0$ by applying $\boldsymbol{H}_0$ to the nature run and adding observation noise, followed by assimilation of $\boldsymbol{y}_0$ using the ETKF to obtain the analysis ensemble $\boldsymbol{X}^a_0$.
5. Propagate $\boldsymbol{X}^a_0$ forward to the forecast time $t = t_f$ to evaluate the forecast ensemble $\boldsymbol{X}^f_{t_f}$.

This procedure is repeated over 10 years (14,600 time steps). The analysis states at $t = 0$ are not recycled into subsequent cycles in order to isolate the contribution of adaptive observations from cycling effects.

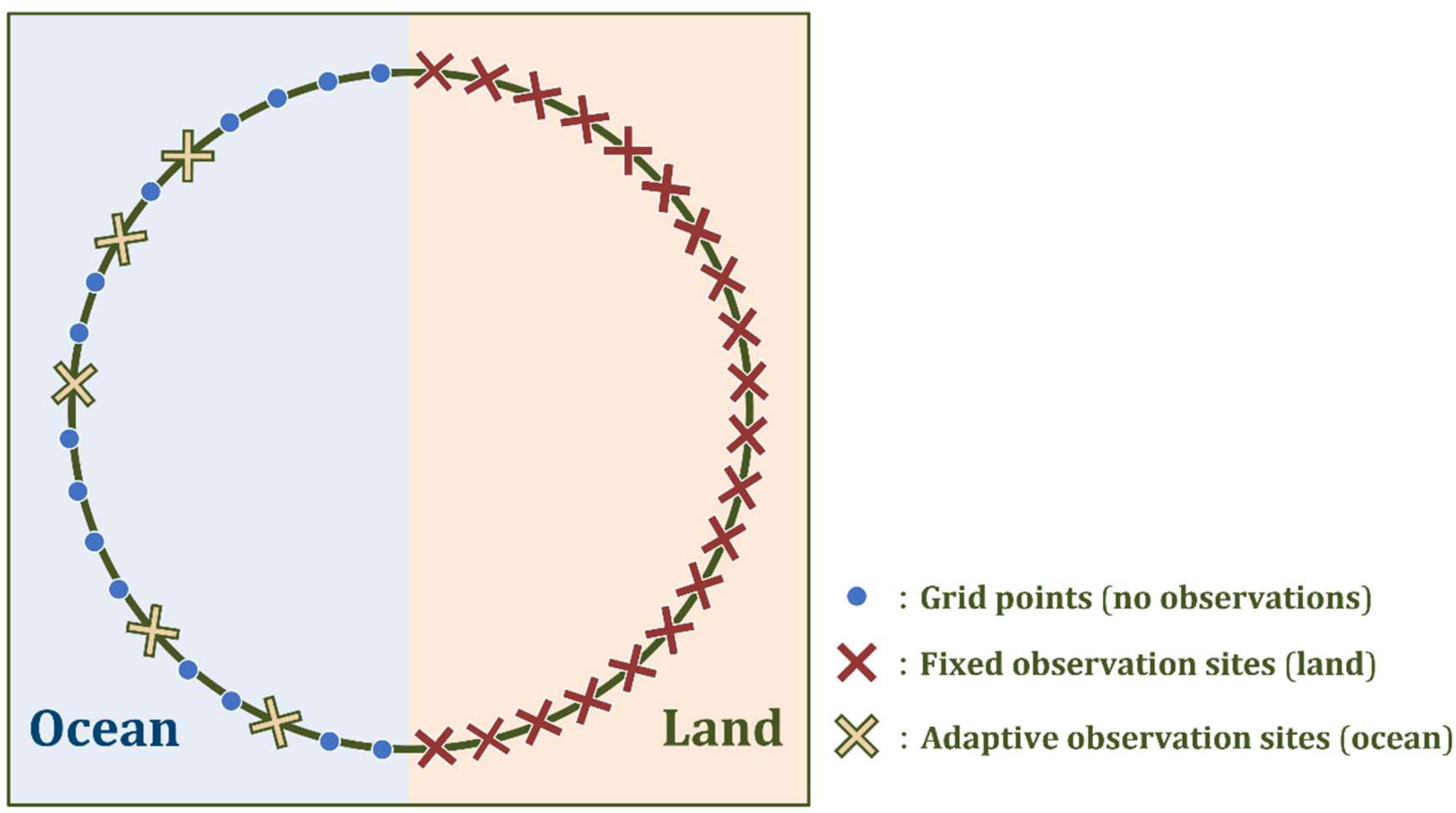

FIGURE 4 Schematic observation layout for the Lorenz-96 model experiments. The 40 grid points are arranged on a circle and divided equally into ocean (left half; blue shading) and land (right half; orange shading). Blue dots denote unobserved grid points. Red crosses indicate fixed observation sites over land, and yellow crosses represent an example of five adaptive observation sites over the ocean.

## 5.2 Spatial Distribution of Selected Observations

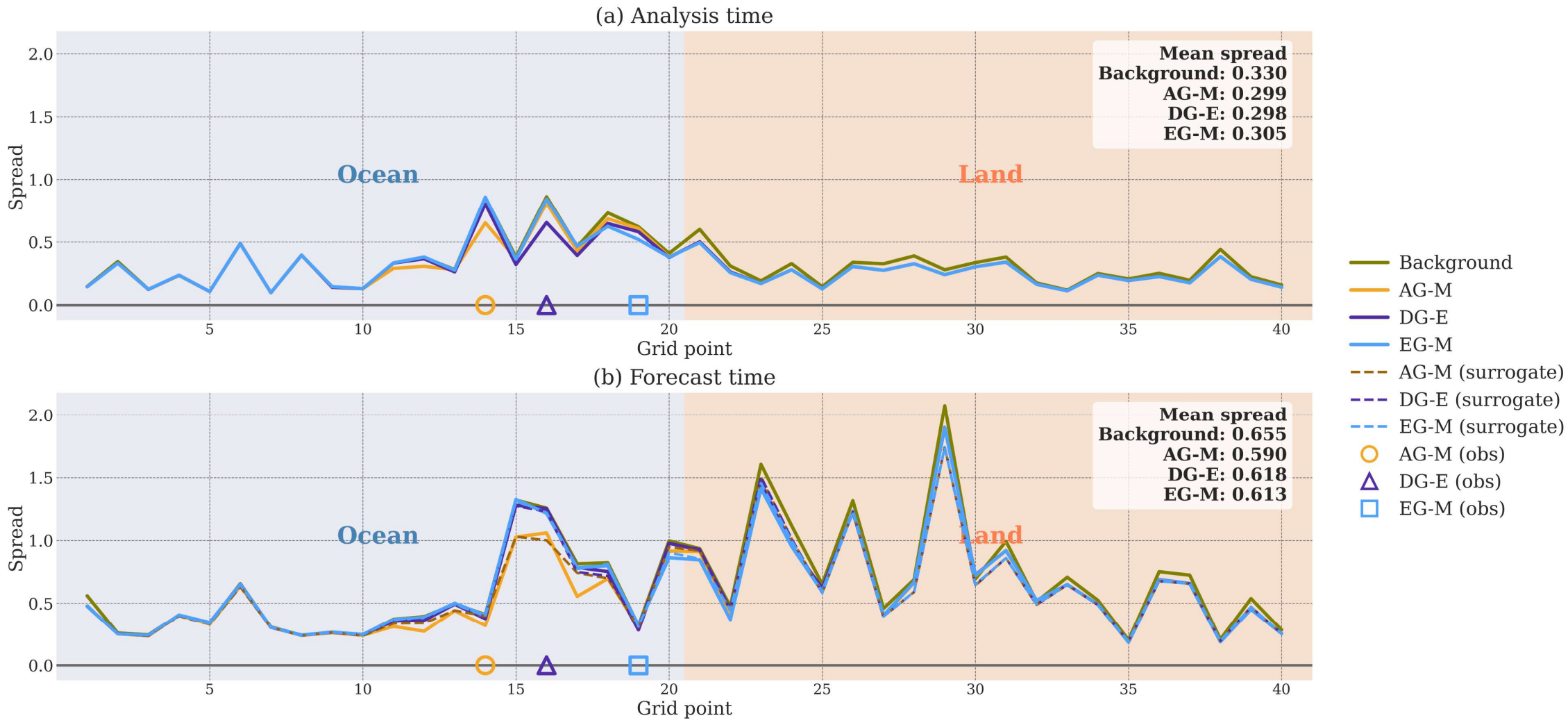

FIGURE 5 Ensemble spreads as functions of grid point for (a) analysis at time $t = 0$ and (b) forecast at time $t = t_f$. Solid lines show the background (green), AG-M (orange), DG-E

(purple), and EG-M (blue) spreads. Dashed lines of the corresponding colors in panel (b) show the forecast spread propagated using the surrogate TLM $\boldsymbol{M}^b$ to validate the surrogate approximation. Symbols indicate the first observation site selected by each method: circle for AG-M, triangle for DG-E, and square for EG-M, respectively. The domain is divided into Ocean (grid points 1–20; blue shading) and Land (grid points 21–40; orange shading). The inset values in each panel indicate the mean spread averaged over all grid points for the background, AG-M, DG-E, and EG-M methods.

Figure 5 shows, at a particular analysis time, the first observation site selected by each greedy method together with the corresponding analysis and forecast ensemble spreads. We focus on the first selected observation site as it is unaffected by previously selected sites, most clearly revealing the characteristic behavior of each objective function. Since ensemble spread is independent of the realized observation values, the differences among AG-M, DG-E, and EG-M reflect purely algorithmic differences. To validate the surrogate TLM $\boldsymbol{M}^b$, the forecast spread propagated using $\boldsymbol{M}^b$ is also shown in Figure 5(b). The two spreads agree closely, confirming that the surrogate approximation is reasonable at the 2-day forecast lead time.

All three methods reduce both the analysis and forecast spreads relative to the background. As expected from the theoretical result in Section 4.2.2, DG-E selects its first observation site at the location of the peak background spread, i.e., the site where the analysis error variance normalized by the observation error variance is largest. While assimilating observations at this site reduces the analysis spread, the forecast spread is larger than those of AG-M and EG-M. This is because DG-E targets analysis error reduction without accounting for the temporal evolution of forecast errors. In contrast, AG-M and EG-M select their first sites away from the background-spread peak, achieving a consistent reduction in forecast spread relative to DG-E. In particular, AG-M achieves the lowest mean forecast spread among the three methods.

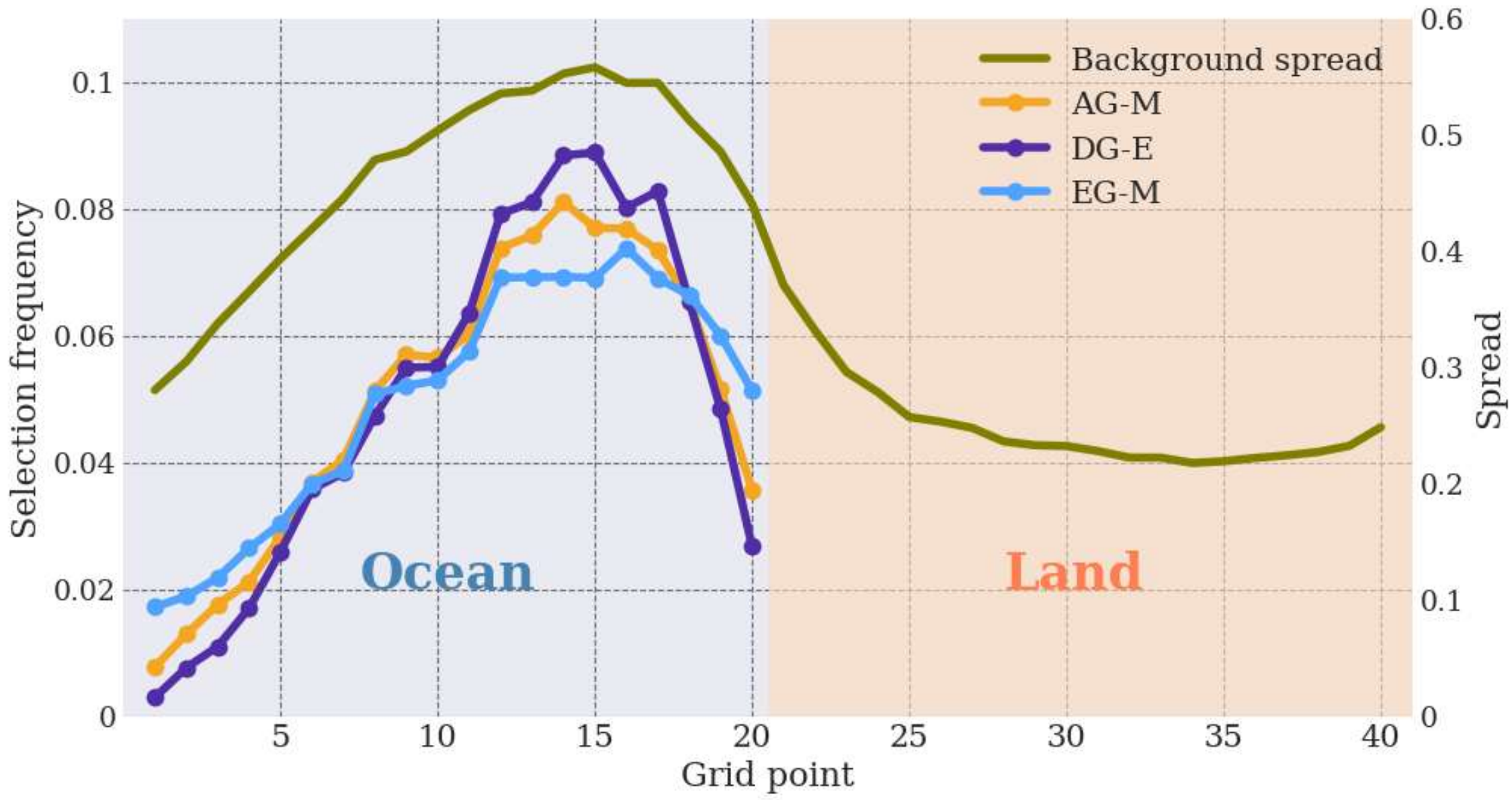

FIGURE 6 Selection frequency of the first adaptive observation site and background ensemble spread as functions of grid point, accumulated over the full 10-year experimental period. Lines show the selection frequency for AG-M (orange), DG-E (purple), and EG-M (blue) on the left axis, and the time-mean background spread (green) on the right axis. The domain is divided into Ocean (grid points 1–20; blue shading) and Land (grid points 21–40; orange shading).

Figure 6 shows the selection frequency of the first observation site together with the time-mean background spread, accumulated over the full 10-year experimental period. In the Lorenz-96 model, errors propagate in the positive grid-point direction (e.g., Kalnay et al., 2012), producing an asymmetric background-spread pattern that is suppressed in the upstream region and larger in the downstream region. Consistent with this pattern, DG-E concentrates its selections near the background-spread peak, whereas AG-M and EG-M select more broadly across off-peak grid points, reflecting their direct targeting of forecast error structures.

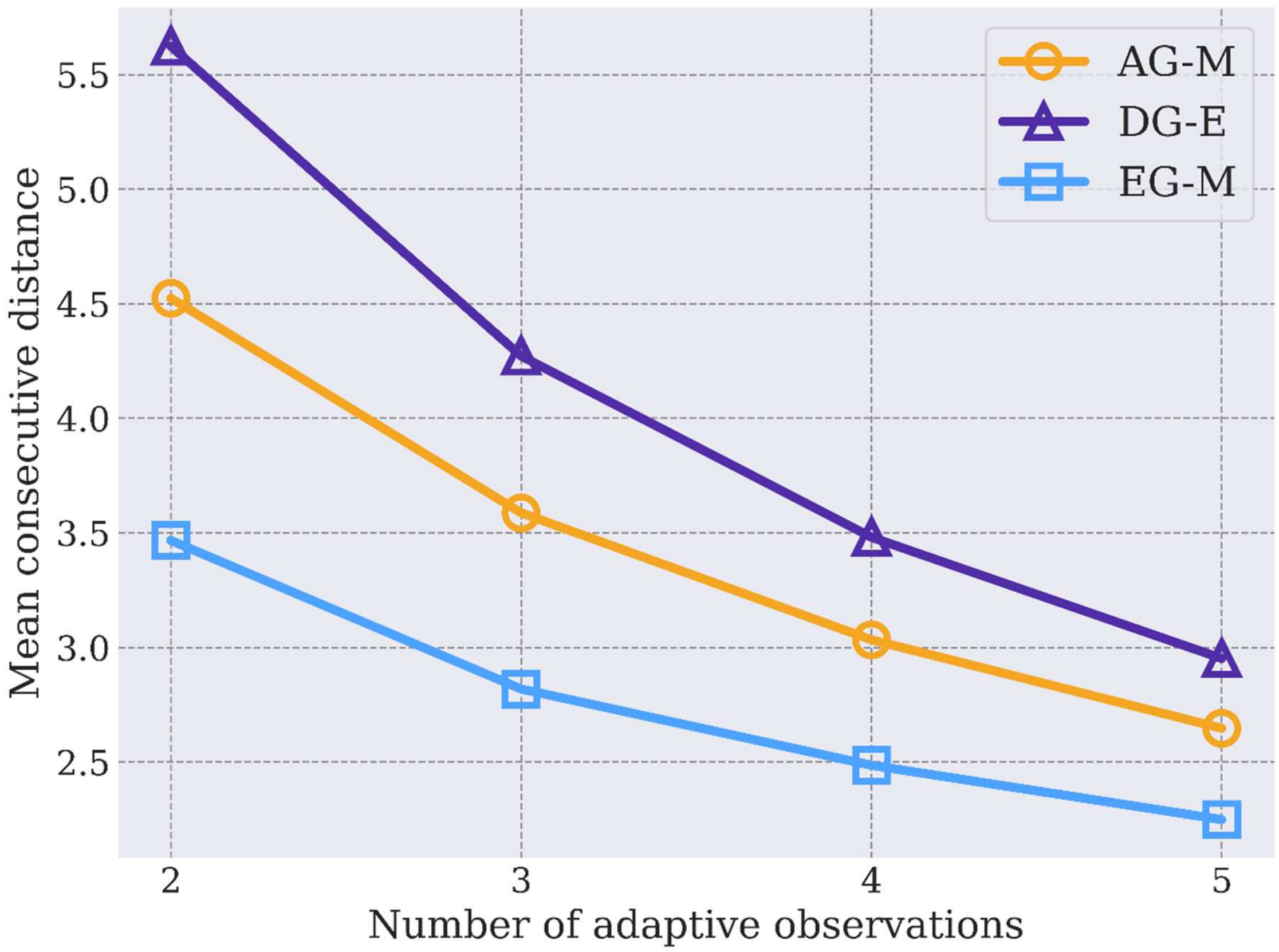

FIGURE 7 Mean distance between consecutively selected adaptive observation sites as a function of the number of adaptive observations, for AG-M (orange circles), DG-E (purple triangles), and EG-M (blue squares). The mean distance is defined as the average of distances between adjacent pairs of selected sites, measured in grid-point units.

Figure 7 shows the mean distance between consecutively selected adaptive observation sites as a function of the number of adaptive observations, where the mean is taken over all adjacent pairs. For example, if five adaptive observation sites are selected at grid points {2, 6, 7, 13, 17}, the consecutive distances are $|6-2|=4$, $|7-6|=1$, $|13-7|=6$, and $|17-13|=4$, giving a mean consecutive distance of $(4+1+6+4)/4=3.75$.

As shown in Figure 7, DG-E yields the largest mean distances, reflecting its tendency to select spatially dispersed sites that gather mutually independent information. In contrast, EG-M produces the smallest mean distances, as it repeatedly targets the dominant error-growth mode, leading to concentrated selections near a single region. Between these two extremes, AG-M yields intermediate configurations.

## 5.3 Forecast Error Evaluation

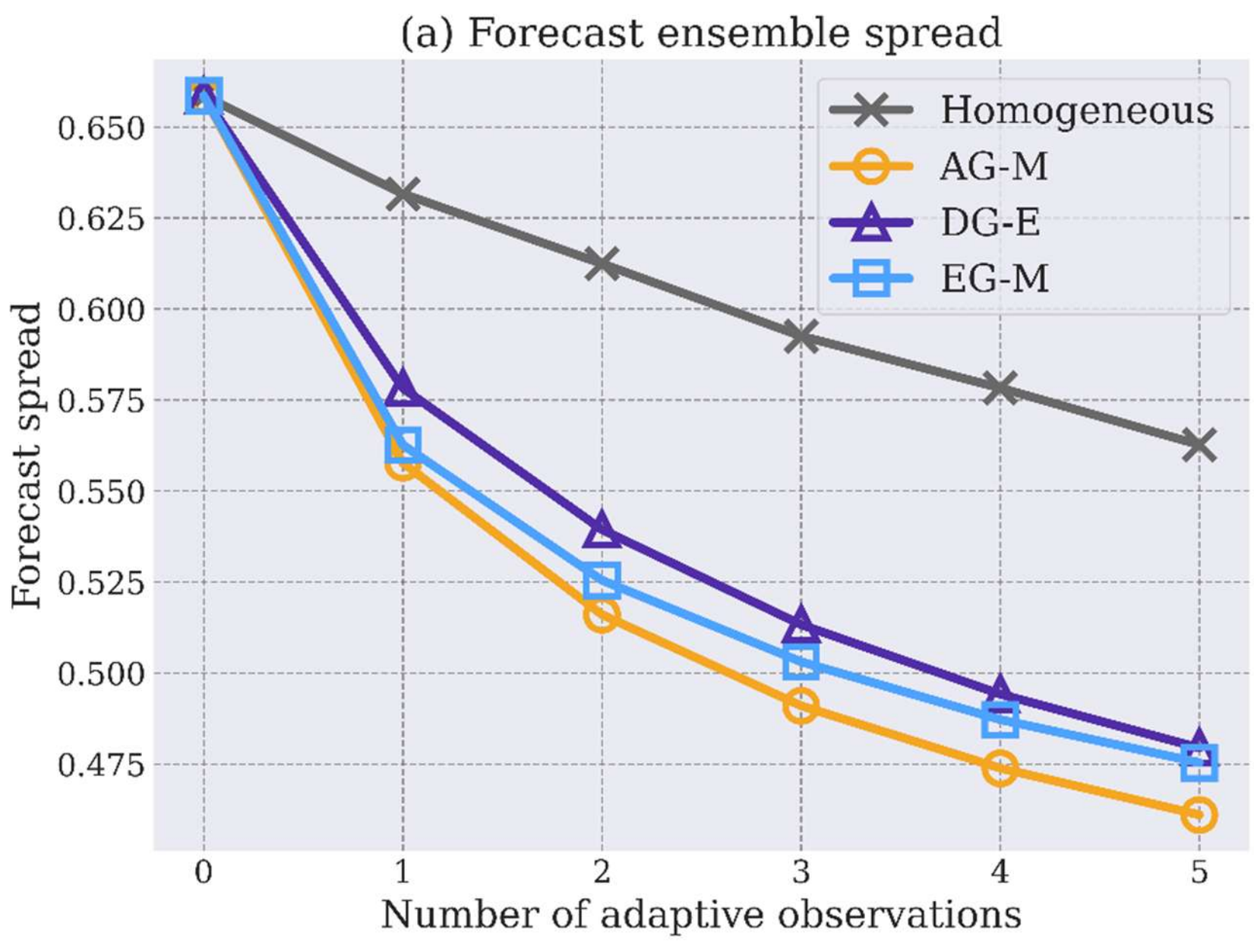

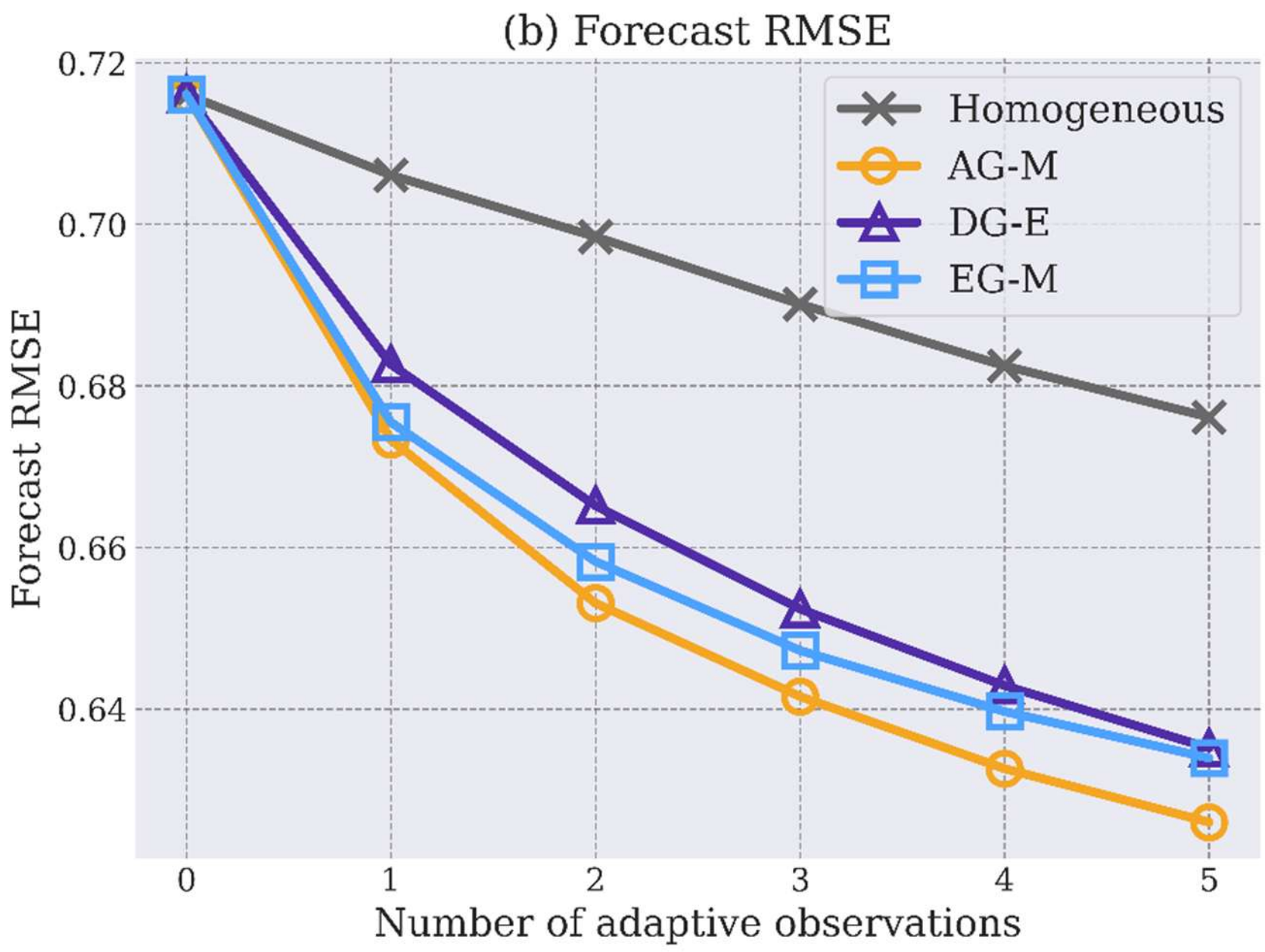

FIGURE 8 (a) Forecast ensemble spread and (b) forecast root-mean-square error (RMSE) as functions of the number of adaptive observations, averaged over the full 10-year experimental period. Lines represent AG-M (orange circles), DG-E (purple triangles), and EG-M (blue squares). Gray lines show the homogeneous configuration (gray crosses) in which adaptive observation sites are placed at equally spaced intervals over the ocean.

Figure 8 shows the forecast ensemble spread and forecast RMSE as functions of the number of adaptive observations. Regardless of the method, optimizing the observation sites substantially reduces both spread and RMSE relative to the homogeneous configuration, demonstrating the effectiveness of greedy selection. Among the optimized methods, AG-M and EG-M, both of which account for forecast error growth through the model-space FIM, outperform DG-E in both metrics. In particular, AG-M yields the smallest spread and RMSE, reflecting its design to minimize the average forecast error variance across all state components. Although EG-M performs comparably to AG-M when the number of adaptive observations is small, the improvement from EG-M saturates as more sites are added, owing to its tendency to concentrate sampling on the dominant error direction.

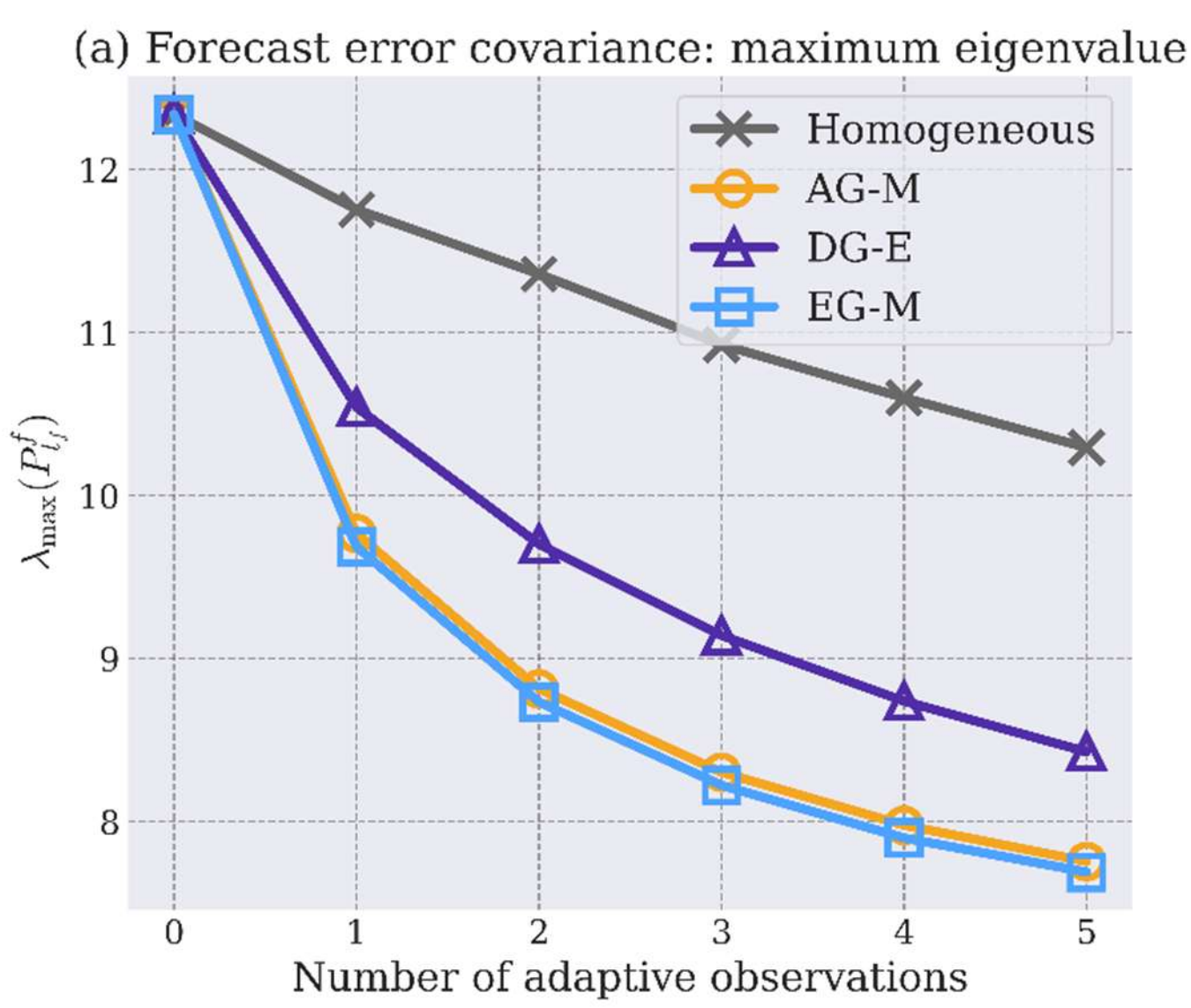

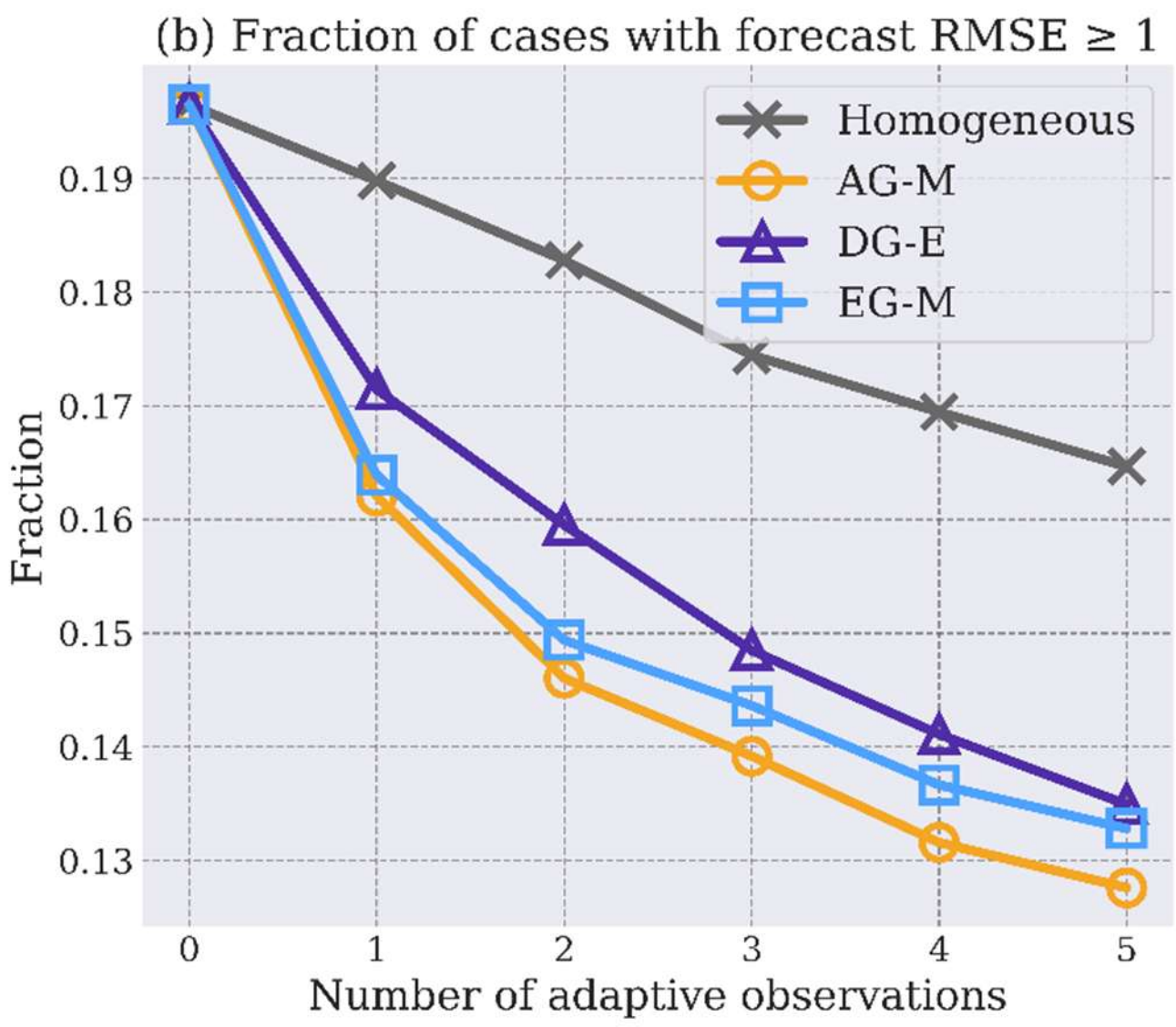

FIGURE 9 (a) Maximum eigenvalue of the forecast error covariance matrix, $\lambda_{\max}\left(\boldsymbol{P}_{t_f}^{f}\right)$, and (b) fraction of cases in which the forecast RMSE exceeds 1.0, as functions of the number of adaptive observations, averaged over the full 10-year experimental period. Lines represent AG-M (orange circles), DG-E (purple triangles), EG-M (blue squares), and the homogeneous configuration (gray crosses).

Figure 9 shows the maximum eigenvalue of the forecast error covariance matrix, $\lambda_{\max}\left(\boldsymbol{P}_{t_f}^{f}\right)$, and the fraction of cases in which the forecast RMSE $\geq 1$. Since EG-M targets the reduction of worst-case forecast errors by minimizing $\lambda_{\max}\left(\boldsymbol{P}_{t_f}^{f}\right)$, these metrics provide a direct assessment of EG-M's target. For $\lambda_{\max}\left(\boldsymbol{P}_{t_f}^{f}\right)$, EG-M yields slightly smaller values than AG-M, consistent with the design of EG-M to directly suppress the dominant error mode, although the difference between the two methods remains small. It is noteworthy that AG-M approaches the performance of EG-M on this metric despite not explicitly optimizing $\lambda_{\max}\left(\boldsymbol{P}_{t_f}^{f}\right)$. For the fraction of cases with forecast RMSE $\geq 1$, however, AG-M outperforms EG-M. This is because EG-M is relatively less effective at reducing error modes other than the dominant one, as it concentrates its selections near a single region, as shown in Figure 7. In contrast, DG-E outperforms the homogeneous configuration but performs worse than AG-M and EG-M on both metrics, consistent with its inability to account for the temporal evolution of forecast errors.

## 5.4 Comparison with Observation Impact Estimation

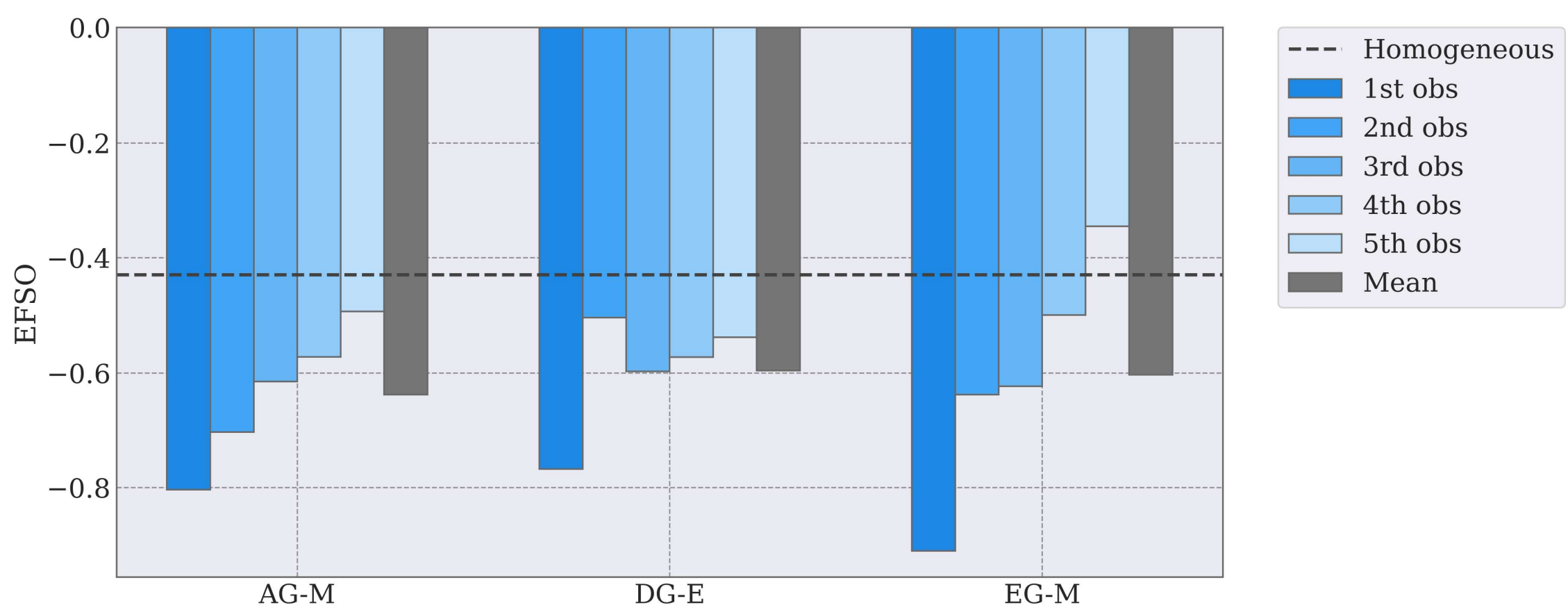

FIGURE 10 Observation impact estimated by the Ensemble Forecast Sensitivity to Observation (EFSO; Kalnay et al., 2012) for five adaptively selected observation sites, for AG-M, DG-E, and EG-M. Colored bars show the EFSO-based forecast impact of each observation in the order selected by the greedy algorithm (1st to 5th), and the gray bar indicates the mean impact across all five observations for each method. The dashed line indicates the mean EFSO of the five adaptively placed observations in the homogeneous configuration. Negative EFSO values indicate a reduction in forecast error. The nature run is used as the verification reference state.

The preceding subsections varied the number of adaptive observations and evaluated forecast metrics. In this subsection, by contrast, we fix the number of adaptive observations at five and decompose the total forecast impact into the contributions of the individual observations, in the order of greedy selection. We quantify this impact using the Ensemble Forecast Sensitivity to Observation (EFSO; Kalnay et al., 2012), which approximates the adjoint-based forecast sensitivity to observation (FSO; Langland and Baker, 2004) using ensemble statistics. The nature run is used as the verification reference state for the EFSO calculation. Although EFSO is a post-assimilation diagnostic and the greedy algorithm relies solely on pre-assimilation background information, consistency between the two provides clear evidence that the greedy selection successfully prioritizes high-impact observations.

Figure 10 shows the EFSO-based forecast impact of each adaptively selected observation in the order of greedy selection. Also shown is the mean EFSO of the five adaptive observations in the homogeneous configuration as a reference baseline. All three methods outperform the homogeneous configuration in terms of mean EFSO, demonstrating the effectiveness of greedy selection. For all methods, the first selected observation exhibits the largest impact. In particular, AG-M achieves the largest mean observation impact among the three methods, and its observation impact decreases approximately monotonically with the order of selection. This behavior indicates the consistency between the AG-M objective of minimizing the mean forecast error variance and the forecast improvement quantified by EFSO. For EG-M, the first selected observation yields the largest impact among the three methods, consistent with its design to preferentially capture the dominant direction of forecast error growth. However, the impacts of subsequent observations decrease rapidly, reflecting the tendency of EG-M to sample redundant error directions. For DG-E, no clear correspondence is observed between selection order and EFSO-based impact, consistent with its theoretical characterization as targeting analysis rather than forecast error reduction.

### 5.5 Computational Cost

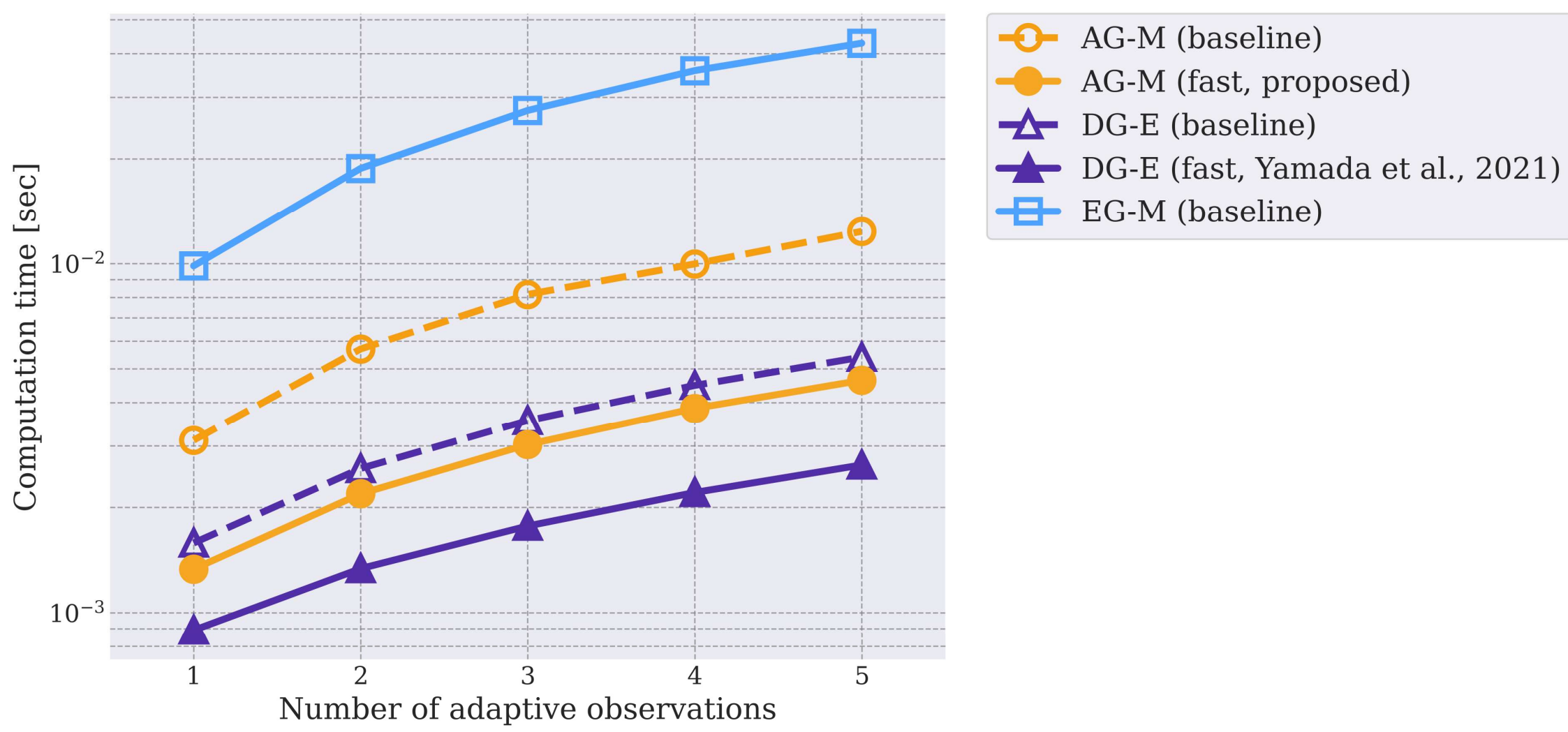

FIGURE 11 Computation time (seconds) as a function of the number of adaptive observations for AG-M baseline (orange open circles), Fast AG-M proposed in this study (orange filled circles), DG-E baseline (purple open triangles), Fast DG-E from Yamada et al. (2021) (purple filled triangles), and EG-M baseline (blue open squares). The vertical axis is shown on a logarithmic scale.

Figure 11 shows the mean computation time for determining observation sites as a function of the number of adaptive observations. Among the methods, Fast DG-E exhibits the shortest computation time, as the determinant computation at each greedy step is replaced by an inexpensive scalar operation, achieving a much shorter computation time than the DG-E baseline. By comparison, Fast AG-M reduces the computation time by approximately a factor of two relative to the baseline AG-M by avoiding the matrix inversion at each greedy step (Section 4.2.1). By contrast, EG-M incurs the highest computation time, as it requires an eigenvalue decomposition at each greedy step. The relative ordering of computation times observed in Figure 11 is consistent with the theoretical complexity analysis summarized in Table 2. Overall, Fast AG-M achieves a computational cost only moderately higher than Fast DG-E, making it a well-balanced method in terms of both forecast accuracy and computational efficiency.

## 6. Summary and Discussion

This study presented a unified theoretical framework integrating SSP and EnKF, and investigated observation network design methods aimed at reducing forecast errors from both theoretical and numerical perspectives. The main theoretical contributions are threefold, connecting SSP and EnKF within optimal experimental design.

First, we formulated the adaptive observation problem for forecast error reduction within the ensemble-based data assimilation cycle. A key consideration in this formulation is the choice of TLM: since the TLM linearized around the analysis trajectory, $\boldsymbol{M}^a$, is unavailable prior to data assimilation, the forecast error covariance must instead be propagated using the surrogate TLM $\boldsymbol{M}^b$, linearized around the background trajectory (see Figure 2). In the ensemble-based framework, $\boldsymbol{M}^b$ is naturally approximated by propagating the background ensemble perturbations to the forecast time. This surrogate approach enables the observation locations to be determined before observation is available, while remaining independent of the choice of observation locations.

Second, we derived the FIM in both the ensemble space and the model space for the EnKF, and clarified the mathematical interpretation of three optimality criteria. The FIM in the ensemble space is defined only at the analysis time and is therefore useful for reducing analysis error, but does not directly characterize forecast error reduction. The FIM in the model space, by contrast, explicitly propagates the error covariance to the forecast time via the TLM, and is therefore the appropriate choice for designing observation networks that directly target forecast error reduction. The three optimality criteria are interpreted accordingly: first, A-optimality in the model space minimizes the mean forecast error variance; second, D-optimality is ill-defined in the model space due to the rank-deficiency of the error covariance matrix and the independence of $\boldsymbol{M}^b$ from observation locations, and is therefore formulated in the ensemble space where it is equivalent to maximizing the Shannon information content of the assimilated observations; and finally, E-optimality in the model space minimizes the largest eigenvalue of the forecast error covariance matrix, corresponding to the worst-case forecast error variance.

Third, we proposed a fast algorithm for AG-M that avoids matrix inversion at each greedy step. This substantially reduces the computational cost relative to the baseline AG-M, as demonstrated in Section 5.5.

A series of numerical experiments using the Lorenz-96 model demonstrated the theoretical findings. Consistent with its rule of selecting the site with the largest analysis error variance, DG-E tended to concentrate its first site near the background error spread peak. In contrast, AG-M and EG-M selected their first sites more broadly, indicating their direct targeting of the forecast error structure. For subsequent selections, DG-E selected more dispersed sites owing to its determinant-based objective, whereas EG-M concentrated its selections near a single region by repeatedly targeting the dominant error-growth mode, with AG-M yielding intermediate configurations between the two. In terms of forecast spread and RMSE, the methods ranked in the order AG-M, EG-M, and DG-E, with all three outperforming the homogeneous baseline. Although EG-M performed comparably to AG-M when the number of adaptive observations was small, its improvement saturated as additional sites were added. For the maximum eigenvalue of the forecast error covariance matrix, EG-M and AG-M yielded nearly indistinguishable values, whereas for the fraction of cases with forecast RMSE exceeding 1.0, AG-M outperformed EG-M. These results are consistent with the theoretical characterization of each method. The validity of the greedy selection approach was further supported by EFSO, which showed that AG-M most consistently prioritizes high-impact observations.

The theoretical and numerical results lead to the conclusion that AG-M is the most consistently reliable method for observation network design aimed at forecast error reduction. It achieves well-balanced reduction of forecast errors across multiple error modes and maintains acceptable computational cost. For worst-case error suppression when only a small number of observations are available, EG-M can be a preferred choice, and DG-E is valuable when the primary objective is to reduce analysis uncertainty.

Several limitations of the present study should be noted. First, we assumed that each observation corresponds to a single physical variable at a single model grid point. The extension to observations that span multiple locations — such as integrated column measurements or remotely sensed retrievals — remains an open question and warrants future investigation. Second, no localization was applied in the present study, whereas operational ensemble data assimilation systems typically employ localization to mitigate sampling errors in large-scale systems. Incorporating localization requires further investigation, as it modifies the structure of the analysis error covariance matrix. One potential approach is ensemble modulation (Bishop et al., 2017), which implements model-space localization within the ETKF while retaining the ensemble-space formulation, and may thus preserve the theoretical properties proved in this study, though this remains to be verified. Finally, additional research should address the behavior of the three criteria under strong non-Gaussianity or nonlinearity; in this regard, approximating the error distributions as Gaussian can yield substantially erroneous estimates of observation impact (Fowler and van Leeuwen, 2012, 2013). Further directions include the influence of model error on observation network design and extensions to multi-time adaptive observation strategies. Such additional work would connect the outcomes of the present study to practical observation network design problems for improving NWP.

ACKNOWLEDGEMENTS:

The authors thank Drs. T. Tsuyuki, K. Okamoto, I. Okabe, and K. Kondo of the Meteorological Research Institute for fruitful discussions. This study was partly supported by the IAAR Research Support Program and VL Program of Chiba University. SK was also supported by the Japan Society for the Promotion of Science (JSPS) KAKENHI grants JP21H04571, JP22K18821, JP25H00752, and the JST Moonshot R&D (JPMJMS2389).

FUNDING INFORMATION:

Japan Society for the Promotion of Science, Grant/Award Numbers: JP21H04571, JP22K18821, JP25H00752; Japan Science and Technology Agency (Moonshot R&D), Grant/Award Number: JPMJMS2389; Chiba University (IAAR Research Support Program and VL Program)

CONFLICT OF INTEREST STATEMENT:

The authors declare no potential conflict of interest.

DATA AVAILABILITY STATEMENT:

The data that support the findings of this study are available from the corresponding author upon reasonable request.

## APPENDIX. Derivation of the fast algorithm for AG in the model space

Consider the situation in which the $k$th observation location is sought in the greedy algorithm. The inverse of the observation error covariance matrix $\left(\boldsymbol{R}_{0,k}^{(i)}\right)^{-1} \in \mathbb{R}^{(N_0+k)\times(N_0+k)}$ for $i \in S_{\mathrm{cand}} \setminus S_{k-1}$ is given by the block matrix inversion formula as

$$\left(\boldsymbol{R}_{0,k}^{(i)}\right)^{-1} = \begin{bmatrix} \boldsymbol{R}_{0,k-1} & \boldsymbol{r}_k^{(i)} \\ \boldsymbol{r}_k^{(i)\top} & \left(\sigma_k^{(i)}\right)^2 \end{bmatrix}^{-1} = \begin{bmatrix} \boldsymbol{R}_{0,k-1}^{-1} + \frac{1}{s_k^{(i)}} \boldsymbol{R}_{0,k-1}^{-1} \boldsymbol{r}_k^{(i)} \boldsymbol{r}_k^{(i)\top} \boldsymbol{R}_{0,k-1}^{-1} & -\frac{1}{s_k^{(i)}} \boldsymbol{R}_{0,k-1}^{-1} \boldsymbol{r}_k^{(i)} \\ -\frac{1}{s_k^{(i)}} \boldsymbol{r}_k^{(i)\top} \boldsymbol{R}_{0,k-1}^{-1} & \frac{1}{s_k^{(i)}} \end{bmatrix}, \quad \text{(A1)}$$

where $s_k^{(i)} = \left(\sigma_k^{(i)}\right)^2 - \boldsymbol{r}_k^{(i)\top} \boldsymbol{R}_{0,k-1}^{-1} \boldsymbol{r}_k^{(i)}$ denotes the Schur complement of $\boldsymbol{R}_{0,k-1}$ in $\boldsymbol{R}_{0,k}^{(i)}$.

Writing the background ensemble perturbations $\delta \boldsymbol{X}_0^b \in \mathbb{R}^{N_{\text{model}} \times N_{\text{ens}}}$ in terms of row vectors $\boldsymbol{w}^{(i)\top} \in \mathbb{R}^{N_{\text{ens}}}$ as in Eq. (25), the inverse of the analysis error covariance matrix in the ensemble space, $\widetilde{\boldsymbol{P}}_{0,k}^{a(i)} \in \mathbb{R}^{N_{\text{ens}} \times N_{\text{ens}}}$, can be expressed as

$$\left(\widetilde{\boldsymbol{P}}_{0,k}^{a(i)}\right)^{-1} = (N_{\text{ens}} - 1)\boldsymbol{I} + \left(\boldsymbol{H}_k^{(i)} \delta \boldsymbol{X}_0^b\right)^\top \left(\boldsymbol{R}_{0,k}^{(i)}\right)^{-1} \left(\boldsymbol{H}_k^{(i)} \delta \boldsymbol{X}_0^b\right). \quad \text{(A2)}$$

Substituting the block structure of $\left(\boldsymbol{R}_{0,k}^{(i)}\right)^{-1}$ and expanding by straightforward algebra yields

$$\left(\widetilde{\boldsymbol{P}}_{0,k}^{a(i)}\right)^{-1} = \left(\widetilde{\boldsymbol{P}}_{0,k-1}^{a}\right)^{-1} + \boldsymbol{d}_k^{(i)} \boldsymbol{d}_k^{(i)\top}, \quad \text{(A3)}$$

where $\boldsymbol{d}_k^{(i)} = \frac{1}{\sqrt{s_k^{(i)}}} \left(\boldsymbol{C}_{k-1}^\top \boldsymbol{R}_{0,k-1}^{-1} \boldsymbol{r}_k^{(i)} - \boldsymbol{w}^{(i)}\right)$. Applying the Sherman–Morrison–Woodbury formula gives

$$\widetilde{\boldsymbol{P}}_{0,k}^{a(i)} = \left[\left(\widetilde{\boldsymbol{P}}_{0,k-1}^{a}\right)^{-1} + \boldsymbol{d}_k^{(i)} \boldsymbol{d}_k^{(i)\top}\right]^{-1} = \widetilde{\boldsymbol{P}}_{0,k-1}^{a} - \frac{\left(\widetilde{\boldsymbol{P}}_{0,k-1}^{a} \boldsymbol{d}_k^{(i)}\right) \left(\widetilde{\boldsymbol{P}}_{0,k-1}^{a} \boldsymbol{d}_k^{(i)}\right)^\top}{1 + \left(\boldsymbol{d}_k^{(i)}\right)^\top \left(\widetilde{\boldsymbol{P}}_{0,k-1}^{a} \boldsymbol{d}_k^{(i)}\right)}. \quad \text{(A4)}$$

The A-optimality criterion in the model space can then be reformulated as

$$\begin{aligned} \arg \min_{i \in S_{\text{cand}} \setminus S_{k-1}} \operatorname{tr}\left(\boldsymbol{P}_{t_f,k}^{f(i)}\right) &= \arg \min_{i \in S_{\text{cand}} \setminus S_{k-1}} \operatorname{tr}\left(\delta \boldsymbol{X}_{t_f}^b \widetilde{\boldsymbol{P}}_{0,k}^{a(i)} \left(\delta \boldsymbol{X}_{t_f}^b\right)^\top\right) \\ &= \arg \max_{i \in S_{\text{cand}} \setminus S_{k-1}} \operatorname{tr}\left(\frac{\left(\widetilde{\boldsymbol{P}}_{0,k-1}^{a} \boldsymbol{d}_k^{(i)}\right) \left(\widetilde{\boldsymbol{P}}_{0,k-1}^{a} \boldsymbol{d}_k^{(i)}\right)^\top}{1 + \left(\boldsymbol{d}_k^{(i)}\right)^\top \left(\widetilde{\boldsymbol{P}}_{0,k-1}^{a} \boldsymbol{d}_k^{(i)}\right)} \left(\delta \boldsymbol{X}_{t_f}^b\right)^\top \delta \boldsymbol{X}_{t_f}^b\right). \end{aligned} \quad \text{(A5)}$$